\documentclass[journal]{IEEEtran}
\usepackage[latin9]{inputenc}
\usepackage{float}
\usepackage{amsmath}
\usepackage{amssymb}
\usepackage{graphicx}

\makeatletter

\providecommand{\tabularnewline}{\\}
\floatstyle{ruled}
\newfloat{algorithm}{tbp}{loa}
\providecommand{\algorithmname}{Algorithm}
\floatname{algorithm}{\protect\algorithmname}

\let\oldforeign@language\foreign@language
\DeclareRobustCommand{\foreign@language}[1]{%
  \lowercase{\oldforeign@language{#1}}}

\usepackage{amsfonts}
\usepackage{mathrsfs}
\usepackage{mathrsfs}\usepackage[noblocks]{authblk}
\usepackage[compress]{cite}
\usepackage{bm}
\usepackage{algorithm}
\usepackage{algorithmic}
\usepackage{epsfig}\usepackage{graphics}\usepackage{subfigure}\usepackage{epsfig}\usepackage{epstopdf}
\usepackage{graphics}\usepackage{subfigure}\usepackage{upgreek}\usepackage{caption}\usepackage{theorem}
\usepackage{breqn}
\usepackage{multicol}
\usepackage{eufrak}
\usepackage{eucal}
\usepackage{array}
\usepackage[compress]{cite}
\usepackage{algorithm}
\usepackage{algorithmic}

\newtheorem{theorem}{Theorem}\newtheorem{lemma}{Lemma}\theoremheaderfont{\normalfont\bfseries}

\makeatother

\begin{document}
\title{Intelligent Reflecting Surface Aided Multigroup Multicast MISO Communication
Systems}
\author{Gui~Zhou,~\IEEEmembership{Student,~IEEE,} Cunhua~Pan,~\IEEEmembership{Member,~IEEE,}
Hong~Ren,~\IEEEmembership{Member,~IEEE,} Kezhi~Wang,~\IEEEmembership{Member,~IEEE,}
and Arumugam~Nallanathan,~\IEEEmembership{Fellow, IEEE}
\thanks{Manuscript received September 26, 2019; revised February 10, 2020 andApril
	6, 2020; accepted April 21, 2020. The associate editor coordinating the review
	of this manuscript and approving it for publication was Prof. Marco Moretti.
	(Corresponding author: Cunhua Pan.)
	
	Gui Zhou, Cunhua Pan, Hong Ren, and Arumugam Nallanathan are with the
	School of Electronic Engineering and Computer Science, Queen Mary University
	of London, London E1 4NS, U.K. (e-mail: g.zhou@qmul.ac.uk; c.pan@
	qmul.ac.uk; h.ren@qmul.ac.uk; a.nallanathan@qmul.ac.uk).
	
	Kezhi Wang is with the Department of Computer and Information
	Q2 Sciences, Northumbria University, Newcastle upon Tyne  NE1 8ST, U.K. (e-mail:
	kezhi.wang@northumbria.ac.uk).
	
	This paper was supported by EP/R006466/1.
	
	Digital Object Identifier 10.1109/TSP.2020.2990098
	 } }
\markboth{Journal of \LaTeX\ Class Files,~Vol.~xx, No.~xx, xx~xxxx}{Zhou \MakeLowercase{\textit{et al.}}: Bare Demo of IEEEtran.cls
for IEEE Journals}
\maketitle
\begin{abstract}
Intelligent reflecting surface (IRS) has recently been envisioned
to offer unprecedented massive multiple-input multiple-output (MIMO)-like
gains by deploying large-scale and low-cost passive reflection elements.
By adjusting the reflection coefficients, the IRS can change the phase
shifts on the impinging electromagnetic waves so that it can smartly
reconfigure the signal propagation environment and enhance the power
of the desired received signal or suppress the interference signal.
In this paper, we consider downlink multigroup multicast communication
systems assisted by an IRS. We aim for maximizing the sum rate of
all the multicasting groups by the joint optimization of the precoding
matrix at the base station (BS) and the reflection coefficients at
the IRS under both the power and unit-modulus constraint. To tackle
this non-convex problem, we propose two efficient algorithms under
the majorization--minimization (MM) algorithm framework. Specifically,
a concave lower bound surrogate objective function of each user's
rate has been derived firstly, based on which two sets of variables
can be updated alternately by solving two corresponding second-order
cone programming (SOCP) problems. Then, in order to reduce the computational
complexity, we derive another concave lower bound function of each
group's rate for each set of variables at every iteration, and obtain
the closed-form solutions under these loose surrogate objective functions.
Finally, the simulation results demonstrate the benefits in terms
of the spectral and energy efficiency of the introduced IRS and the
effectiveness in terms of the convergence and complexity of our proposed
algorithms. 
\end{abstract}


\IEEEpeerreviewmaketitle{}

\section{Introduction}

In the era of 5G and Internet of Things by 2020, it is predicted that
the network capacity will increase by 1000 folds to serve at least
50 billions devices through wireless communications \cite{IMT-vision2015}
and the capacity is expected to be achieved with lower energy consumption.
To meet those Quality of Service (QoS) requirements, intelligent reflecting
surface (IRS), as a promising new technology, has been proposed recently
to achieve high spectral and energy efficiency. It is an artificial
passive radio array structure where the phase of each passive element
on the surface can be adjusted continuously or discretely with low
power consumption \cite{cui2014coding,liu2019intelligent}, and then
change the directions of the reflected signal into the specific receivers
to enhance the received signal power \cite{qingqing2019,Pan2019multicell,Pan2019intelleget,Chongwen2020JSAC}
or suppress interference as well as enhance security/privacy \cite{Xianghao2009,AN2020}.

The IRS, as a new concept beyond conventional massive multiple-input
and multiple-output (MIMO) systems, maintains all the advantages of
massive MIMO systems, such as being capable of focusing large amounts
of energy in three-dimensional space which paves the way for wireless
charging, remote sensing and data transmissions. However, the differences
between IRS and massive MIMO are also obvious. Firstly, the IRS can
be densely deployed in indoor spaces, making it possible to provide
high data rates for indoor devices in the way of near-field communications
\cite{Tan2016}. Secondly, in contrast to conventional active antenna
array equipped with energy-consuming radio frequency chains and power
amplifiers, the IRS with passive reflection elements is cost-effective
and energy-efficient \cite{qingqing2019}, which enables IRS to be
a prospective energy-efficient technology in green communications.
Thirdly, as the IRS just reflects the signal in a passive way, there
is no thermal noise or self-interference imposed on the received signal
as in conventional full-duplex relays.

Due to these significant advantages, IRS has been investigated in
various wireless communication systems. Specifically, the authors
in \cite{qingqing2019} first formulated the joint active and passive
beamforming design problem both in downlink single-user and multiple-users
multiple-input single-output (MISO) systems assisted by the IRS, while
the total transmit power of the base station (BS) is minimized based
on the semidefinite relaxation (SDR) \cite{luo2010SDR} and alternating
optimization (AO) techniques. In order to reduce the high computational
complexity incurred by SDR, Yu \textit{et al.} proposed low complexity
algorithms based on MM (Majorization--Minimization or Minorization--Maximization)
algorithm in \cite{Xianghao2009} and manifold optimization in \cite{yu2019miso}
to design reflection coefficients with the targets of maximizing the
security capacity and spectral efficiency communications, respectively.
Pan \textit{et al. }considered the weighted sum rate maximization
problems in multicell MIMO communications \cite{Pan2019multicell},
simultaneous wireless information and power transfer (SWIPT) aided
systems \cite{Pan2019intelleget}, artificial-noise-aided secure MIMO
communications \cite{AN2020}, all demonstrating the significant performance
gains achieved by deploying an IRS in the networks. A deep reinforcement
learning (DRL)-based algorithm \cite{Chongwen2020JSAC} and a mobile
edge computing-based algorithm \cite{baitong2019} were proposed to
jointly design the active and passive beamformings in IRS-related
systems. In cognitive radio (CR) communication systems, the high rate
for the secondary user (SU) can be acheived with the assistance of
the IRS \cite{sharing2020letter}.

However, all the above-mentioned contributions only investigated the
performance benefits of deploying an IRS in unicast transmissions,
where the BS sends an independent data stream to each user. However,
unicast transmissions will cause severe interference and high system
complexity when the number of users is large. To address this issue,
the multicast transmission based on content reuse \cite{Golrezaei2013}
(e.g., identical content may be requested by a group of users simultaneously)
has attracted wide attention, especially for the application scenarios
such as popular TV programme or video conference. From the perspective
of operators, it can be envisioned that multicast transmission is
capable of effectively alleviating the pressure of tremendous wireless
data traffic and play a vital role in the next generation wireless
networks. Therefore, it is necessary to explore the potential performance
benefits brought by an IRS during the multigroup multicast transmission.
In specifically, in multicast systems, the data rate of each group
is limited by the user with the worst-channel gains. Hence, the IRS
can be deployed to improve the channel conditions of the worst-case
user, which can be significantly improve the system performance.

A common performance metric in multicast transmissions is the max-min
fairness (MMF), where the minimum signal-to-interference-plus-noise-ratio
(SINR) or spectral efficiency of users in each multicasting group
or among all multicasting groups is maximized \cite{Luo2006transmit,luo2008quality,Tran2014conic,Xiang2014massive,Sadeghi2017reducing}.
Prior seminal treatments of multicast transmission in single-group
and multigroup are presented in \cite{Luo2006transmit,luo2008quality},
where the MMF problems are formulated as a fractional second-order
cone programming (SOCP) and are NP-hard in general. The SDR technique
\cite{luo2010SDR} was adopted to approximately solve the SOCP problem
with some mathematical manipulations. In order to reduce the high
computational complexity of SDR, several low-complexity algorithms,
such as successive convex approximation approach in the single-group
multicast scenario \cite{Tran2014conic}, asymptotic approach \cite{Xiang2014massive}
and heuristic algorithm \cite{Sadeghi2017reducing} in the multigroup
multicast scenario, have been proposed by exploiting the special feature
of near-orthogonal massive MIMO channels.

In this paper, we consider an IRS-assisted multigroup multicast transmission
system in which a multiple-antenna BS transmits independent information
data streams to multiple groups, and the single-antenna users in the
same group share the same information and suffer from interference
from those signals sent to other groups. Unfortunately, the popular
SDR-based method incurs a high computational complexity which hinders
its practical implementation when the number of design parameters
(e.g., precoding matrix and reflection coefficient vector) becomes
large. Furthermore, the aforementioned low-complexity techniques designed
for the IRS-aided unicast communication schemes cannot be directly
applied in the multigroup multicast communication systems since the
MMF metric is a non-differentiable and complex objective function.

Against the above background, the main contributions of our work are
summarized as follows: 
\begin{itemize}
\item To the best of our knowledge, this is the first work exploring the
performance benefits of deploying an IRS in multigroup multicast communication
systems. Specifically, we jointly optimize the precoding matrix and
the reflection coefficient vector to maximize the sum rate of all
the multicasting groups, where the rate of each multicasting group
is limited by the minimum rate of users in the group. This formulated
problem is much more challenging than previous problems considered
in unicast systems since our considered objective function is non-differentiable
and complex due to the nature of the multicast transmission mechanism.
In addition, the highly coupled variables and complex sum rate expression
aggravates the difficulty to solve this problem. 
\item The formulated problem is solved efficiently in an iterative manner
based on the alternating optimization method under the MM algorithm
framework. Specifically, we firstly minorize the original non-concave
objective function by a surrogate function which is biconcave of precoding
matrix and reflection coefficient vector, and then apply the alternating
optimization method to decouple those variables. At each iteration
of the alternating optimization method, the subproblem corresponding
to each set of variables is reformulated as an SOCP problem by introducing
auxiliary variables, which can help to transform the non-differentiable
concave objective function into a series of convex constraints. 
\item To further reduce the computational complexity, we use the MM method
to derive closed-form solutions of each subproblem, instead of solving
the complex SOCP problems with a high complexity at each iteration.
Specifically, we firstly apply the log-sum-exp lower bound to approximate
the non-differentiable concave objective function, yielding a differentiable
concave function. Then, we derive a tractable surrogate objective
function of the log-sum-exp function, based on which we derive the
closed-form solutions of each subproblem. Finally, we prove that the
proposed algorithm is guaranteed to converge and the solution sequences
generated by the algorithm converge to KKT points. 
\item Finally, the simulation results demonstrate the superiority of the
IRS-assisted multigroup multicast system over conventional massive
MIMO systems in terms of the spectral efficiency and energy efficiency.
The convergence and the low complexity of the proposed algorithms
have also been illustrated. 
\end{itemize}
The remainder of this paper is organized as follows. Section II introduces
the system model and formulates the optimization problem. An SOCP-based
method is developed to solve the problem in Section III. Section IV
further provides a low-complexity algorithm. Finally, Section V and
Section VI show the simulation results and conclusions, respectively.

\noindent \textbf{Notations:} The following mathematical notations
and symbols are used throughout this paper. Vectors and matrices are
denoted by boldface lowercase letters and boldface uppercase letters,
respectively. The symbols $\mathbf{X}^{*}$, $\mathbf{X}^{\mathrm{T}}$,
$\mathbf{X}^{\mathrm{H}}$, and $||\mathbf{X}||_{F}$ denote the conjugate,
transpose, Hermitian (conjugate transpose), Frobenius norm of matrix
$\mathbf{X}$, respectively. The symbols $||\mathbf{x}||_{1}$ and
$||\mathbf{x}||_{2}$ denote 1-norm and 2-norm of vector $\mathbf{x}$,
respectively. The symbols $\mathrm{Tr}\{\cdot\}$, $\mathrm{Re}\{\cdot\}$,
$|\cdot|$, and $\angle\left(\cdot\right)$ denote the trace, real
part, modulus and angle of a complex number, respectively. $\mathrm{diag}(\mathbf{x})$
is a diagonal matrix with the entries of $\mathbf{x}$ on its main
diagonal. $[\mathbf{x}]_{m}$ means the $m^{\mathrm{th}}$ element
of the vector $\mathbf{x}$. The Kronecker product between two matrices
$\mathbf{X}$ and $\mathbf{Y}$ is denoted by $\mathbf{X}\otimes\mathbf{Y}$.
$\mathbf{X}\succeq\mathbf{Y}$means that $\mathbf{X}-\mathbf{Y}$
is positive semidefinite. Additionally, the symbol $\mathbb{C}$ denotes
complex field, $\mathbb{R}$ represents real field, and $j\triangleq\sqrt{-1}$
is the imaginary unit.

\section{System Model}

\subsection{Signal Transmission Model}

\begin{figure}
\centering \includegraphics[width=3.6in,height=2.7in]{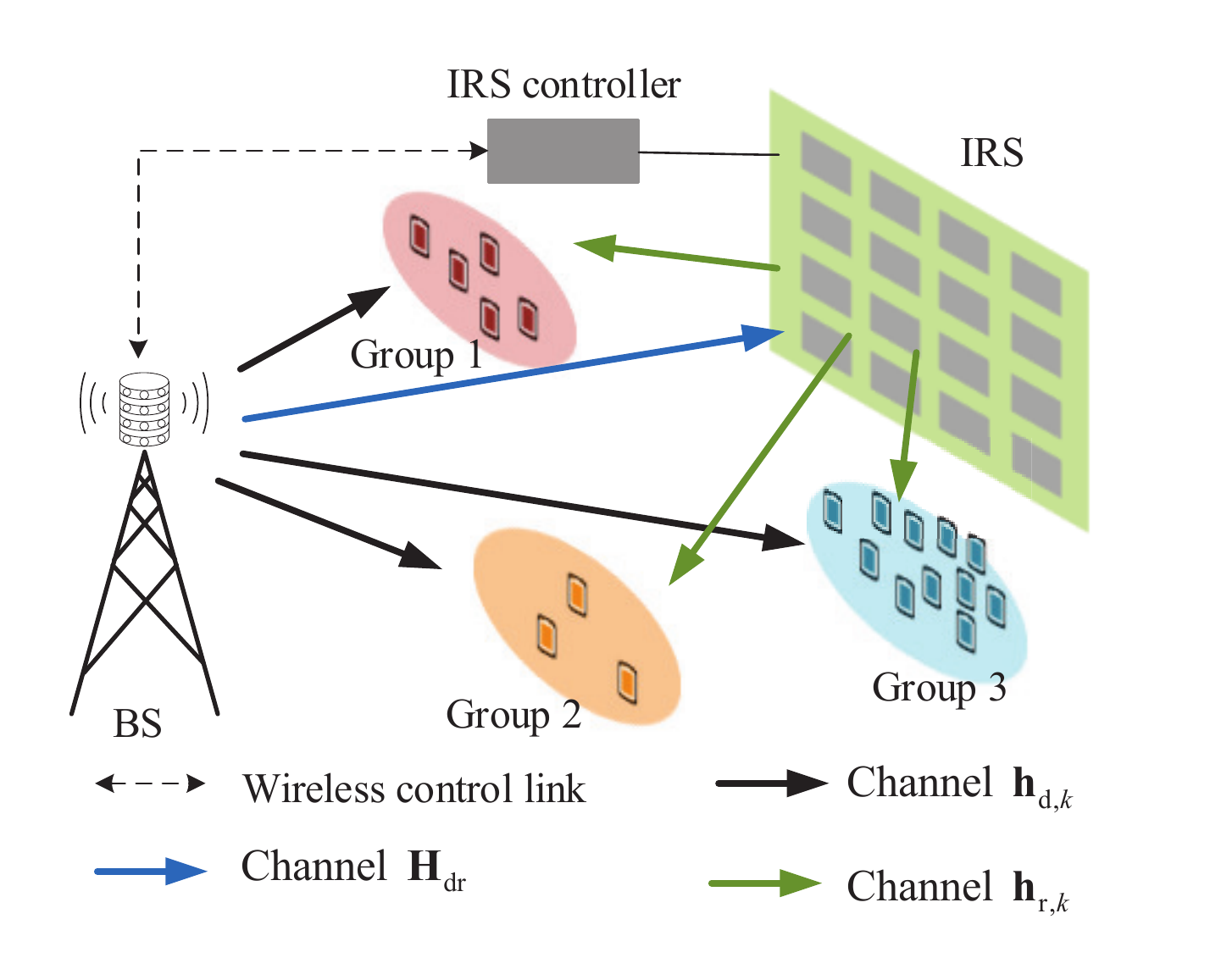} \caption{An IRS-aided multigroup multicast communication system.}
\label{system-model} 
\end{figure}

As shown in Fig. \ref{system-model}, we consider an IRS-aided multigroup
multicast MISO communication system. There is a BS with $N$ transmit
antennas serving $G$ multicasting groups. Users in the same group
share the same information data and the information data destined
for different groups are independent and different, which means there
exists inter-group interference. Let us define the set of all multicast
groups by $\mathcal{G}=\{1,2,...,G\}$. Assuming that there are $K(K\geq G)$
users in total, the user set belonging to group $g\in\mathcal{G}$
is denoted as $\mathcal{K}_{g}$ and each user can only belong to
one group, i.e., $\mathcal{K}_{i}$$\cap$$\mathcal{K}_{j}$=$\emptyset,\forall i,j\in\mathcal{G},i\neq j$.
The transmit signal at the BS is 
\begin{equation}
{\bf x}=\sum\limits _{g=1}^{G}{\bf f}_{g}s_{g},
\end{equation}
where $s_{g}$ is the desired independent Gaussian data symbol of
group $g$ and follows $\mathbb{E}[|s_{g}|^{2}]=1$ as well as ${\bf f}_{g}\in\mathbb{C}^{N\times1}$
is the corresponding precoding vector. Let us denote the collection
of all precoding vectors as $\mathbf{F=}[{\bf \mathbf{f}}_{1},\cdots,{\bf \mathbf{f}}_{G}]\in\mathbb{C}^{N\times G}$
satisfying the power constraint $\mathcal{S}_{F}=\{\mathbf{F}\mid\mathrm{Tr}\left[\mathbf{F}^{\mathrm{H}}\mathbf{F}\right]\leq P_{\mathrm{T}}\}$,
where $P_{\mathrm{T}}$ is the maximum available transmit power at
the BS.

In the multigroup multicast system, we propose to employ an IRS with
the goal of enhancing the received signal strength of users by reflecting
signals from the BS to the users. It is assumed that the signal power
of the multi-reflections (i.e., reflections more than once) on the
IRS is ignored due to the severe path loss \cite{qingqing2019}. Denote
$M$ as the number of the reflection elements on the IRS, then the
reflection coefficient matrix of the IRS is modeled by a diagonal
matrix $\mathbf{E}=\mathrm{diag}([e_{1},\cdots,e_{M}]^{\mathrm{T}})\in\mathbb{C}^{M\times M}$,
where $|e_{m}|^{2}=1,\forall m=1,\cdots,M$ \cite{qingqing2019}.
Please note that the design of the practical reflection amplitude
which was modeled as a function of the phase shifts \cite{Samith2020}
is more complex and will be investigated in our future work. The channels
spanning from the BS to user $k$, from the BS to the IRS, and from
the IRS to user $k$ are denoted by $\mathbf{h}_{\mathrm{d},k}\in\mathbb{C}^{N\times1}$,
$\mathbf{H_{\mathrm{dr}}}\in\mathbb{C}^{M\times N}$, and $\mathbf{h}_{\mathrm{r},k}\in\mathbb{C}^{M\times1}$,
respectively.

It is assumed that the channel state information (CSI) is perfectly
known at the BS. The BS is responsible for designing the reflection
coefficients of the IRS and sends them back to the IRS controller
as shown in Fig. \ref{system-model}. As a result, the received signal
of user $k\in\mathcal{K}_{g}$ belonging to group $g$ is 
\begin{equation}
y_{k}=(\mathbf{h}_{\mathrm{d},k}^{\mathrm{H}}+\mathbf{h}_{\mathrm{r},k}^{\mathrm{H}}\mathbf{E}\mathbf{H_{\mathrm{dr}}})\sum\limits _{g=1}^{G}{\bf f}_{g}s_{g}+n_{k},
\end{equation}
where $n_{k}$ is the received noise at user $k$, which is an additive
white Gaussian noise (AWGN) following circularly symmetric complex
Gaussian (CSCG) distribution with zero mean and variance $\sigma_{k}^{2}$.
Then, its achievable data rate (bps/Hz) is given by 
\begin{equation}
R_{k}=\log_{2}\left(1+\frac{|(\mathbf{h}_{\mathrm{d},k}^{\mathrm{H}}+\mathbf{h}_{\mathrm{r},k}^{\mathrm{H}}\mathbf{E}\mathbf{H_{\mathrm{dr}}}){\bf f}_{g}|^{2}}{\sum_{i\neq g}^{G}|(\mathbf{h}_{\mathrm{d},k}^{\mathrm{H}}+\mathbf{h}_{\mathrm{r},k}^{\mathrm{H}}\mathbf{E}\mathbf{H_{\mathrm{dr}}}){\bf f}_{i}|^{2}+\sigma_{k}^{2}}\right).\label{eq:rate-1}
\end{equation}

Denoting by $\mathbf{H}_{k}=\left[\begin{array}{c}
\mathrm{diag}(\mathbf{h}_{\mathrm{r},k}^{\mathrm{H}})\mathbf{H_{\mathrm{dr}}}\\
\mathbf{h}_{\mathrm{d},k}^{\mathrm{H}}
\end{array}\right]\in\mathbb{C}^{(M+1)\times N}$ the equivalent channel spanning from the BS to user $k$ and by $\mathbf{e}=[e_{1},\cdots e_{M},1]^{\mathrm{T}}\in\mathbb{C}^{(M+1)\times1}$
the equivalent reflection coefficient vector, we have 
\begin{align}
|(\mathbf{h}_{\mathrm{d},k}^{\mathrm{H}}+\mathbf{h}_{\mathrm{r},k}^{\mathrm{H}}\mathbf{E}\mathbf{H_{\mathrm{dr}}}){\bf f}_{g}|^{2} & =|\mathbf{e}^{\mathrm{H}}\mathbf{H}_{k}{\bf f}_{g}|^{2},\\
\sum_{i\neq g}^{G}|(\mathbf{h}_{\mathrm{d},k}^{\mathrm{H}}+\mathbf{h}_{\mathrm{r},k}^{\mathrm{H}}\mathbf{E}\mathbf{H_{\mathrm{dr}}}){\bf f}_{i}|^{2} & =\sum_{i\neq g}^{G}|\mathbf{e}^{\mathrm{H}}\mathbf{H}_{k}{\bf f}_{i}|^{2}+\sigma_{k}^{2}.
\end{align}
Note that $\mathbf{e}$ belongs to the set $\mathcal{S}_{e}=\{\mathbf{e}\mid|e_{m}|^{2}=1,1\leq m\leq M,e_{M+1}=1\}$.
Then, the data rate expression in (\ref{eq:rate-1}) can be rewritten
in a compact form as 
\begin{equation}
R_{k}\left(\mathbf{F},\mathbf{e}\right)=\log_{2}\left(1+\frac{|\mathbf{e}^{\mathrm{H}}\mathbf{H}_{k}{\bf f}_{g}|^{2}}{\sum_{i\neq g}^{G}|\mathbf{e}^{\mathrm{H}}\mathbf{H}_{k}{\bf f}_{i}|^{2}+\sigma_{k}^{2}}\right).\label{eq:Rate-k-1}
\end{equation}

Due to the nature of the multicast mechanism, the achievable data
rate of group $g$ is limited by the minimum user rate in this group
and is defined as follows 
\begin{equation}
\min_{k\in\mathcal{K}_{g}}\left\{ R_{k}\left(\mathbf{F},\mathbf{e}\right)\right\} .
\end{equation}

\subsection{Problem Formulation}

In this paper, we aim to jointly optimize the precoding matrix $\mathbf{F}$
and reflection coefficient vector $\mathbf{e}$ to maximize the sum
rate of the whole system, which is defined as the sum rate achieved
by all groups. Mathematically, the optimization problem is formulated
as 
\begin{align}
\mathop{\max}\limits _{\mathbf{F},\mathbf{e}} & \;\;\left\{ F\left(\mathbf{F},\mathbf{e}\right)=\sum_{g=1}^{G}\min_{k\in\mathcal{K}_{g}}\left\{ R_{k}\left(\mathbf{F},\mathbf{e}\right)\right\} \right\} \nonumber \\
{\rm s.t.} & \thinspace\thinspace\thinspace\mathbf{F}\in\mathcal{S}_{F},\mathbf{e}\in\mathcal{S}_{e}.\label{eq:Problem-original}
\end{align}

Problem (\ref{eq:Problem-original}) is a non-convex problem and difficult
to solve since the objective function $F\left(\mathbf{F},\mathbf{e}\right)$
is non-differentiable and non-concave, while the unit-modulus constraint
set $\mathcal{S}_{e}$ is also non-convex. In the following, we propose
two efficient algorithms based on the MM algorithm framework to solve
Problem (\ref{eq:Problem-original}).

\subsection{Majorization-Minimization Method}

The aim of the MM method \cite{Hunter2004MM,MM} is to find an easy-to-solve
surrogate problem with a surrogate objective function, then optimize
it instead of the original complex one. Specifically, suppose that
$f(\mathbf{x})$ is the original objective function which needs to
be maximized over a convex set $\mathcal{S}_{x}$. Let $\widetilde{f}(\mathbf{x}|\mathbf{x}^{n})$
denote a real-valued function of variable $\mathbf{x}$ with given
$\mathbf{x}^{n}$. The function $\widetilde{f}(\mathbf{x}|\mathbf{x}^{n})$
is said to minorize $f(\mathbf{x})$ at a given point $\mathbf{x}^{n}$
if they satisfy the following conditions \cite{MM}: 
\begin{align*}
\mathrm{(A1):} & \widetilde{f}(\mathbf{x}^{n}|\mathbf{x}^{n})=f(\mathbf{x}^{n}),\forall\mathbf{x}^{n}\in\mathcal{S}_{x};\\
\mathrm{(A2):} & \widetilde{f}(\mathbf{x}|\mathbf{x}^{n})\leq f(\mathbf{x}),\forall\mathbf{x},\mathbf{x}^{n}\in\mathcal{S}_{x};\\
\mathrm{(A3):} & \widetilde{f}^{'}(\mathbf{x}|\mathbf{x}^{n};\mathbf{d})|_{\mathbf{x}=\mathbf{x}^{n}}=f^{'}(\mathbf{x}^{n};\mathbf{d}),\forall\mathbf{d}\thinspace\thinspace\mathrm{\textrm{with}}\thinspace\thinspace\mathbf{x}^{n}+\mathbf{d}\in\mathcal{S}_{x};\\
\mathrm{(A4):} & \widetilde{f}(\mathbf{x}|\mathbf{x}^{n})\thinspace\thinspace\textrm{is continuous in \ensuremath{\mathbf{x}} and \ensuremath{\mathbf{x}^{n}}. }
\end{align*}
where $f^{'}(\mathbf{x}^{n};\mathbf{d})$, defined as the direction
derivative of $f(\mathbf{x}^{n})$ in the direction $\mathbf{d}$,
is 
\[
f^{'}(\mathbf{x}^{n};\mathbf{d})=\underset{\lambda\rightarrow0}{\lim}\frac{f(\mathbf{x}^{n}+\lambda\mathbf{d})-f(\mathbf{x}^{n})}{\lambda}.
\]

\section{SOCP-based MM method}

In this section, we propose an SOCP-based MM method to solve Problem
(8). Specifically, under the MM algorithm framework, we first handle
the non-convex objective function by introducing its concave surrogate
function. Then, we adopt the alternating optimization method to solve
the subproblems corresponding to different sets of variables alternately.

Note that $F\left(\mathbf{F},\mathbf{e}\right)$ is a composite function
which is the linear combinations of some pointwise minimum with non-concave
subfunction $R_{k}\left(\mathbf{F},\mathbf{e}\right)$. We first tackle
the non-concave property of $R_{k}\left(\mathbf{F},\mathbf{e}\right)$.
To this end, we introduce the following lemma.

\begin{lemma}\label{lemma-1}

Let $\{\mathbf{F}^{n},\mathbf{e}^{n}\}$ be the solutions obtained
at iteration $n-1$, then $R_{k}\left(\mathbf{F},\mathbf{e}\right)$
is minorized by a concave surrogate function $\widetilde{R}_{k}\left(\mathbf{F},\mathbf{e}|\mathbf{F}^{n},\mathbf{e}^{n}\right)$
defined by 
\begin{align}
 & \widetilde{R}_{k}\left(\mathbf{F},\mathbf{e}|\mathbf{F}^{n},\mathbf{e}^{n}\right)\nonumber \\
 & =\textrm{const}_{k}+2\mathrm{Re}\left\{ a_{k}\mathbf{e}^{\mathrm{H}}\mathbf{H}_{k}{\bf f}_{g}\right\} -b_{k}\sum_{i=1}^{G}|\mathbf{e}^{\mathrm{H}}\mathbf{H}_{k}{\bf f}_{i}|^{2}\nonumber \\
 & \leq R_{k}\left(\mathbf{F},\mathbf{e}\right),\label{eq:Rate-surrogate}
\end{align}
where 
\begin{align*}
 & a_{k}=\frac{({\bf f}_{g}^{n})^{\mathrm{H}}\mathbf{H}_{k}^{\mathrm{H}}\mathbf{e}^{n}}{\sum_{i\neq g}^{G}|(\mathbf{e}^{n})^{\mathrm{H}}\mathbf{H}_{k}{\bf f}_{i}^{n}|^{2}+\sigma_{k}^{2}},\\
 & b_{k}=\frac{|(\mathbf{e}^{n})^{\mathrm{H}}\mathbf{H}_{k}{\bf f}_{g}^{n}|^{2}}{\left(\sum_{i\neq g}^{G}|(\mathbf{e}^{n})^{\mathrm{H}}\mathbf{H}_{k}{\bf f}_{i}^{n}|^{2}+\sigma_{k}^{2}\right)\left(\sum_{i=1}^{G}|(\mathbf{e}^{n})^{\mathrm{H}}\mathbf{H}_{k}{\bf f}_{i}^{n}|^{2}+\sigma_{k}^{2}\right)},\\
 & \textrm{const}_{k}=R_{k}\left(\mathbf{F}^{n},\mathbf{e}^{n}\right)-b_{k}\sigma_{k}^{2}-b_{k}\left(\sum_{i=1}^{G}|(\mathbf{e}^{n})^{\mathrm{H}}\mathbf{H}_{k}{\bf f}_{i}^{n}|^{2}+\sigma_{k}^{2}\right),
\end{align*}
at fixed point $\{\mathbf{F}^{n},\mathbf{e}^{n}\}$.

\end{lemma}

\textbf{\textit{Proof: }}Please refer to Appendix \ref{subsec:The-proof-of-1}.\hspace{3cm}$\blacksquare$

Based on the above theorem, Problem (\ref{eq:Problem-original}) can
be transformed into the following surrogate problem: 
\begin{align}
\mathop{\max}\limits _{\mathbf{F},\mathbf{e}} & \;\;\left\{ \widetilde{F}\left(\mathbf{F},\mathbf{e}|\mathbf{F}^{n},\mathbf{e}^{n}\right)=\sum_{g=1}^{G}\min_{k\in\mathcal{K}_{g}}\left\{ \widetilde{R}_{k}\left(\mathbf{F},\mathbf{e}|\mathbf{F}^{n},\mathbf{e}^{n}\right)\right\} \right\} \nonumber \\
{\rm s.t.} & \thinspace\thinspace\thinspace\thinspace\mathbf{F}\in\mathcal{S}_{F},\mathbf{e}\in\mathcal{S}_{e}.\label{eq:Problem-MM}
\end{align}

We note that $\widetilde{R}_{k}\left(\mathbf{F},\mathbf{e}|\mathbf{F}^{n},\mathbf{e}^{n}\right)$
is biconcave of $\mathbf{F}$ and $\mathbf{e}$ \cite{gorski2007biconvex},
since $\widetilde{R}_{k}\left(\mathbf{F}|\mathbf{F}^{n}\right)=\widetilde{R}_{k}\left(\mathbf{F},\mathbf{e}|\mathbf{F}^{n},\mathbf{e}^{n}\right)$
with given $\mathbf{e}$ is concave of $\mathbf{F}$ and $\widetilde{R}_{k}\left(\mathbf{e}|\mathbf{e}^{n}\right)=\widetilde{R}_{k}\left(\mathbf{F},\mathbf{e}|\mathbf{F}^{n},\mathbf{e}^{n}\right)$
with given $\mathbf{F}$ is concave of $\mathbf{e}$. This biconvex
problem enables us to use the alternating optimization (AO) method
to alternately update $\mathbf{F}$ and $\mathbf{e}$.

\subsection{Optimizing the Precoding Matrix $\mathbf{F}$ }

In this subsection, we aim to optimize the precoding matrix $\mathbf{F}$
with given $\mathbf{e}$. With some manipulations, $\widetilde{R}_{k}\left(\mathbf{F},\mathbf{e}|\mathbf{F}^{n},\mathbf{e}^{n}\right)$
in (\ref{eq:Rate-surrogate}) can be shown to be a quadratic function
of $\mathbf{F}$: 
\begin{align}
\widetilde{R}_{k}\left(\mathbf{F}|\mathbf{F}^{n}\right) & =\textrm{const}_{k}+2\textrm{\ensuremath{\mathrm{Re}}}\left\{ a_{k}\mathbf{e}^{\mathrm{H}}\mathbf{H}_{k}{\bf f}_{g}\right\} -b_{k}\sum_{i=1}^{G}|\mathbf{e}^{\mathrm{H}}\mathbf{H}_{k}{\bf f}_{i}|^{2}\nonumber \\
 & =\textrm{const}_{k}+2\textrm{\ensuremath{\mathrm{Re}}}\left\{ \mathrm{Tr}\left[\mathbf{C}_{k}^{\mathrm{H}}\mathbf{F}\right]\right\} -\mathrm{Tr}\left[\mathbf{F}^{\mathrm{H}}\mathbf{B}_{k}\mathbf{F}\right],\label{eq:Rate-surrogate-F}
\end{align}
where $\mathbf{B}_{k}=b_{k}\mathbf{H}_{k}^{\mathrm{H}}\mathbf{e}\mathbf{e}^{\mathrm{H}}\mathbf{H}_{k}$,
$\mathbf{C}_{k}^{\mathrm{H}}=a_{k}\mathbf{t}_{g}\mathbf{e}^{\mathrm{H}}\mathbf{H}_{k}$,
and $\mathbf{t}_{g}\in\mathbb{R}^{G\times1}$ is a selection vector
in which the $g^{\mathrm{th}}$ element is equal to one and all the
other elements are equal to zero.

By using (\ref{eq:Rate-surrogate-F}), the subproblem of Problem (\ref{eq:Problem-MM})
for the optimization of $\mathbf{F}$ is 
\begin{align}
\mathop{\max}\limits _{\mathbf{F}} & \;\;\sum_{g=1}^{G}\min_{k\in\mathcal{K}_{g}}\left\{ \textrm{const}_{k}+2\textrm{\ensuremath{\mathrm{Re}}}\left\{ \mathrm{Tr}\left[\mathbf{C}_{k}^{\mathrm{H}}\mathbf{F}\right]\right\} -\mathrm{Tr}\left[\mathbf{F}^{\mathrm{H}}\mathbf{B}_{k}\mathbf{F}\right]\right\} \nonumber \\
{\rm s.t.} & \thinspace\thinspace\thinspace\mathbf{F}\in\mathcal{S}_{F}.\label{eq:Problem-F-1}
\end{align}
We then tackle the pointwise minimum expressions in the objective
function of Problem (\ref{eq:Problem-F-1}) by introducing auxiliary
variables $\boldsymbol{\gamma}=\text{[}\gamma_{1},...,\gamma_{G}]^{\mathrm{T}}$,
as follows 
\begin{align}
\mathop{\max}\limits _{\mathbf{F},\boldsymbol{\gamma}} & \;\;\sum_{g=1}^{G}\gamma_{g}\nonumber \\
{\rm s.t.} & \thinspace\thinspace\thinspace\thinspace\mathbf{F}\in\mathcal{S}_{F},\nonumber \\
 & \thinspace\thinspace\thinspace\thinspace\textrm{const}_{k}+2\textrm{\ensuremath{\mathrm{Re}}}\left\{ \mathrm{Tr}\left[\mathbf{C}_{k}^{\mathrm{H}}\mathbf{F}\right]\right\} -\mathrm{Tr}\left[\mathbf{F}^{\mathrm{H}}\mathbf{B}_{k}\mathbf{F}\right]\geq\gamma_{g},\nonumber \\
 & \thinspace\thinspace\thinspace\thinspace\thinspace\forall k\in\mathcal{K}_{g},\forall g\in\mathcal{G}.\label{eq:Problem-F-socp}
\end{align}
Problem (\ref{eq:Problem-F-socp}) is an SOCP problem and the globally
solution can be obtained by the CVX \cite{CVX2018} solver, such as
MOSEK \cite{Mosek2018}.

\subsection{Optimizing the reflection coefficient vector $\mathbf{e}$}

In this subsection, we focus on optimizing the reflection coefficient
vector $\mathbf{e}$ with given $\mathbf{F}$, then $\widetilde{R}_{k}\left(\mathbf{e}|\mathbf{e}^{n}\right)$
can be rewritten as 
\begin{align}
\widetilde{R}_{k}\left(\mathbf{e}|\mathbf{e}^{n}\right) & =\textrm{const}_{k}+2\textrm{\textrm{\ensuremath{\mathrm{Re}}}}\left\{ \mathbf{a}_{k}^{\mathrm{H}}\mathbf{e}\right\} -\mathbf{e}^{\mathrm{H}}\mathbf{A}_{k}\mathbf{e},\label{eq:rate-e}
\end{align}
where $\mathbf{A}_{k}=b_{k}\mathbf{H}_{k}\sum_{i=1}^{G}{\bf f}_{i}{\bf f}_{i}^{\mathrm{H}}\mathbf{H}_{k}^{\mathrm{H}}$
and $\mathbf{a}_{k}=a_{k}\mathbf{H}_{k}{\bf f}_{g}$.

Upon replacing the objective function of Problem (\ref{eq:Problem-MM})
by (\ref{eq:rate-e}), the subproblem for the optimization of $\mathbf{e}$
is given by 
\begin{align}
\mathop{\max}\limits _{\mathbf{e}} & \;\;\sum_{g=1}^{G}\min_{k\in\mathcal{K}_{g}}\left\{ \textrm{const}_{k}+2\textrm{\ensuremath{\mathrm{Re}}}\left\{ \mathbf{a}_{k}^{\mathrm{H}}\mathbf{e}\right\} -\mathbf{e}^{\mathrm{H}}\mathbf{A}_{k}\mathbf{e}\right\} \nonumber \\
{\rm s.t.} & \thinspace\thinspace\thinspace\thinspace\mathbf{e}\in\mathcal{S}_{e}.\label{eq:Problem-e-1}
\end{align}

Also introducing auxiliary variables $\boldsymbol{\kappa}=\text{[}\kappa_{1},...,\kappa_{G}]^{\mathrm{T}}$,
Problem (\ref{eq:Problem-e-1}) is equivalent to 
\begin{align}
\mathop{\max}\limits _{\mathbf{e},\boldsymbol{\kappa}} & \;\;\sum_{g=1}^{G}\kappa_{g}\nonumber \\
{\rm s.t.} & \thinspace\thinspace\thinspace\thinspace\mathbf{e}\in\mathcal{S}_{e},\nonumber \\
 & \thinspace\thinspace\thinspace\thinspace\textrm{const}_{k}+2\textrm{\ensuremath{\mathrm{Re}}}\left\{ \mathbf{a}_{k}^{\mathrm{H}}\mathbf{e}\right\} -\mathbf{e}^{\mathrm{H}}\mathbf{A}_{k}\mathbf{e}\geq\kappa_{g},\nonumber \\
 & \thinspace\thinspace\thinspace\thinspace\forall k\in\mathcal{K}_{g},\forall g\in\mathcal{G}.\label{eq:Problem-e-2}
\end{align}

The above problem is still non-convex due to the non-convex unit-modulus
set $\mathcal{S}_{e}$. To address this issue, we replace it with
a relaxed convex one as 
\[
\mathcal{S}_{e-relax}=\{\mathbf{e}^{\mathrm{H}}\textrm{diag}(\mathbf{i}_{m})\mathbf{e}\leq1,\;\;\forall m=1,\cdots,M,e_{M+1}=1\},
\]
where $\mathbf{i}_{m}\in\mathbb{R}^{(M+1)\times1}$ is a selection
vector whose $m^{\mathrm{th}}$ element is equal to one and all the
other elements are equal to zero. Let us denote by $\widehat{\mathbf{e}}_{1}$
the optimal solution of the following relaxed version of the SOCP
problem, i.e., 
\begin{align}
\widehat{\mathbf{e}}_{1}=\mathrm{arg}\mathop{\max}\limits _{\mathbf{e}} & \;\;\sum_{g=1}^{G}\gamma_{g}\nonumber \\
{\rm s.t.} & \thinspace\thinspace\thinspace\thinspace\mathbf{e}\in\mathcal{S}_{e-relax},\nonumber \\
 & \thinspace\thinspace\thinspace\thinspace\textrm{const}_{k}+2\textrm{\ensuremath{\mathrm{Re}}}\left\{ \mathbf{a}_{k}^{\mathrm{H}}\mathbf{e}\right\} -\mathbf{e}^{\mathrm{H}}\mathbf{A}_{k}\mathbf{e}\geq\kappa_{g},\nonumber \\
 & \thinspace\thinspace\thinspace\thinspace\forall k\in\mathcal{K}_{g},\forall g\in\mathcal{G}.\label{eq:Problem-e-SOCP}
\end{align}
Then, the locally optimal solution $\mathbf{e}$ in the $n^{th}$
iteration is 
\begin{equation}
\mathbf{e}^{n+1}=\begin{cases}
\widehat{\mathbf{e}}_{2}\text{,} & \textrm{if }F\left(\mathbf{F}^{n+1},\widehat{\mathbf{e}}_{2}|\mathbf{F}^{n},\mathbf{e}^{n}\right)\geq F\left(\mathbf{F}^{n+1},\mathbf{e}^{n}|\mathbf{F}^{n},\mathbf{e}^{n}\right),\\
\mathbf{e}^{n}, & \textrm{otherwise},
\end{cases}\label{eq:dfd}
\end{equation}
where 
\begin{equation}
\widehat{\mathbf{e}}_{2}=\exp\left\{ j\angle\left(\frac{\widehat{\mathbf{e}}_{1}}{\left[\widehat{\mathbf{e}}_{1}\right]_{M+1}}\right)\right\} ,
\end{equation}
and symbol $\left[\widehat{\mathbf{e}}_{1}\right]_{m}$ denotes the
$m^{\mathrm{th}}$ element of the vector $\widehat{\mathbf{e}}_{1}$.
Here the $\exp\left\{ \cdot\right\} $ and the $\angle\left(\cdot\right)$
are both element-wise operations.

\subsection{Algorithm development}

Based on the above analysis, Algorithm \ref{Algorithm-SOCP} summarizes
the alternating update process between precoding matrix $\mathbf{F}$
and reflection coefficient vector $\mathbf{e}$ to maximize the sum
rate of the whole system.

\begin{algorithm}
\caption{SOCP-based MM algorithm}
\label{Algorithm-SOCP} \begin{algorithmic}[1] \REQUIRE Initialize
$\mathbf{F}^{0}$ and $\mathbf{e}^{0}$, and $n=0$.

\REPEAT

\STATE Calculate $\mathbf{F}^{n+1}$ by solving Problem (\ref{eq:Problem-F-socp})
with given $\mathbf{e}^{n}$;

\STATE Calculate $\mathbf{e}^{n+1}$ by solving Problem (\ref{eq:Problem-e-SOCP})
with given $\mathbf{F}^{n+1}$;

\STATE $n\leftarrow n+1$;

\UNTIL The value of function $F\left(\mathbf{F},\mathbf{e}\right)$
in (\ref{eq:Problem-original}) converges. \end{algorithmic} 
\end{algorithm}

\subsubsection{Complexity analysis}

Now we analyze the computational complexity of Algorithm \ref{Algorithm-SOCP},
which mainly comes from optimizing $\mathbf{F}$ in the SOCP problem
in (\ref{eq:Problem-F-socp}) and optimizing $\mathbf{e}$ in the
SOCP problem in (\ref{eq:Problem-e-SOCP}).

According to \cite{Ben-Tal2001convex}, the complexity of solving
an SOCP problem, with $M_{\mathrm{socp}}$ second order cone constraints
where the dimension of each is $N_{\mathrm{socp}}$, is $\mathcal{O}(N_{\mathrm{socp}}M_{\mathrm{socp}}^{3.5}+N_{\mathrm{socp}}^{3}M_{\mathrm{socp}}^{2.5})$.
Problem (\ref{eq:Problem-F-socp}) contains one power constraint with
dimension $NG$ and $K$ rate constraints with dimension $NG$. Therefore,
the complexity of solving Problem (\ref{eq:Problem-F-socp}) per iteration
is $\mathcal{O}(NG+N^{3}G^{3}+NGK^{3.5}+N^{3}G^{3}K^{2.5})$. Problem
(\ref{eq:Problem-e-SOCP}) has $M$ constant modulus constraints with
dimension one for sparse vector $\mathbf{i}_{m}$ and $K$ rate constraints
with dimension $M+1$. Therefore, the complexity of solving Problem
(\ref{eq:Problem-e-SOCP}) per iteration is $\mathcal{O}(M^{3.5}+M^{2.5}+(M+1)K^{3.5}+(M+1)^{3}K^{2.5})$.
Therefore, the approximate complexity of Algorithm \ref{Algorithm-SOCP}
per iteration is $\mathcal{O}(N^{3}G^{3}K^{2.5}+M^{3.5}+MK^{3.5})$.

\subsubsection{Convergence analysis}

The following theorem shows the convergence and solution properties
of Algorithm \ref{Algorithm-SOCP}.

\begin{theorem}\label{Theorem-socp-kkt} The objective function value
sequence $\{F\left(\mathbf{F}^{n},\mathbf{e}^{n}\right)\}$ generated
by Algorithm \ref{Algorithm-SOCP} is guaranteed to converge, and
the optimal solution converges to a Karush-Kuhn-Tucker (KKT) point.

\end{theorem}

\textbf{\textit{Proof: }}Please refer to Appendix \ref{subsec:The-proof-of-5}.\hspace{3cm}$\blacksquare$

\section{Low-complexity MM method}

As seen in Algorithm \ref{Algorithm-SOCP}, we need to solve two SOCP
problems in each iteration, which incurs a high computational complexity.
In this section, we aim to derive a low-complexity algorithm containing
closed-form solutions.

Since $\min_{k\in\mathcal{K}_{g}}\left\{ \widetilde{R}_{k}\left(\mathbf{F},\mathbf{e}|\mathbf{F}^{n},\mathbf{e}^{n}\right)\right\} $
in Problem (\ref{eq:Problem-MM}) is non-differentiable, we approximate
it as a smooth function by using the following smooth log--sum--exp
lower-bound \cite{xu2001smoothing} 
\begin{align}
 & \min_{k\in\mathcal{K}_{g}}\left\{ \widetilde{R}_{k}\left(\mathbf{F},\mathbf{e}|\mathbf{F}^{n},\mathbf{e}^{n}\right)\right\} \approx f_{g}\left(\mathbf{F},\mathbf{e}\right)\nonumber \\
 & =-\frac{1}{\mu_{g}}\log\Bigl(\sum_{k\in\mathcal{K}_{g}}\mathrm{exp}\left\{ -\mu_{g}\widetilde{R}_{k}\left(\mathbf{F},\mathbf{e}|\mathbf{F}^{n},\mathbf{e}^{n}\right)\right\} \Bigr),
\end{align}
where $\mu_{g}>0$ is a smoothing parameter which satisfies 
\begin{align}
f_{g}\left(\mathbf{F},\mathbf{e}\right) & \leq\min_{k\in\mathcal{K}_{g}}\left\{ \widetilde{R}_{k}\left(\mathbf{F},\mathbf{e}|\mathbf{F}^{n},\mathbf{e}^{n}\right)\right\} \nonumber \\
 & \leq f_{g}\left(\mathbf{F},\mathbf{e}\right)+\frac{1}{\mu_{g}}\log\left(|\mathcal{K}_{g}|\right).
\end{align}

\begin{theorem}

$f_{g}\left(\mathbf{F},\mathbf{e}\right)$ is biconcave of $\mathbf{F}$
and $\mathbf{e}$.

\end{theorem}

\textbf{\textit{Proof: }}According to \cite{book-convex}, if the
Hessian matrix of a function is semi-negative definite, that function
is concave. In particular, we derive the Hessian matrix of the exp-sum-log
function $f\left(x\right)=-\log\Bigl(\sum_{k\in\mathcal{K}_{g}}\mathrm{exp}\left\{ -x\right\} \Bigr)$
as 
\begin{equation}
\nabla^{2}f(x)=-\frac{1}{\left(\mathbf{1}\mathbf{z}^{\mathrm{T}}\right)^{2}}\left(\left(\mathbf{1}^{\mathrm{T}}\mathbf{z}\right)\mathrm{diag}(\mathbf{z})-\mathbf{z}\mathbf{z}^{\mathrm{T}}\right),\label{eq:Hessian}
\end{equation}
where $\mathbf{z}=(e^{x_{1}},\ldots,e^{x_{N}})$. Then for all $\mathbf{v}$,
we have 
\begin{align}
 & \mathbf{v}^{\mathrm{T}}\nabla^{2}f(x)\mathbf{v}\nonumber \\
 & =-\frac{1}{\left(\mathbf{1}\mathbf{z}^{\mathrm{T}}\right)^{2}}\left(\left(\sum_{n=1}^{N}z_{n}\right)\left(\sum_{n=1}^{N}v_{n}^{2}z_{n}\right)-\left(\sum_{n=1}^{N}v_{n}z_{n}\right)^{2}\right)\nonumber \\
 & =-\left(\mathbf{b}^{\mathrm{T}}\mathbf{b}\mathbf{a}^{\mathrm{T}}\mathbf{a}-\left(\mathbf{a}^{\mathrm{T}}\mathbf{b}\right)^{2}\right)\leq0,
\end{align}
where the components of vectors $\mathbf{a}$ and $\mathbf{b}$ are
$a_{n}=v_{n}\sqrt{z_{n}}$ and $b_{n}=\sqrt{z_{n}}$, respectively.
The inequality follows from the Cauchy-Schwarz inequality. Then $\nabla^{2}f(x)\preceq0$,
and the log-sum-exp function $f\left(x\right)$ is concave. Therefore,
$-\frac{1}{\mu_{g}}\log\left(\sum_{k\in\mathcal{K}_{g}}\mathrm{exp}\left\{ -\mu_{g}\widetilde{R}_{k}\right\} \right)$
is an increasing and concave function w.r.t. $\widetilde{R}_{k}$.
Recall that $\widetilde{R}_{k}\left(\mathbf{F},\mathbf{e}|\mathbf{F}^{n},\mathbf{e}^{n}\right)$
is biconcave of $\mathbf{F}$ and $\mathbf{e}$. Finally, according
to the composition principle \cite{book-convex}, $f_{g}\left(\mathbf{F},\mathbf{e}\right)$
is biconcave of $\mathbf{F}$ and $\mathbf{e}$. The proof is complete.
\hspace{7cm}$\blacksquare$

Large $\mu_{g}$ leads to high accuracy of the approximation, but
it also causes the problem to be nearly ill-conditioned. When $\mu_{g}$
is chosen appropriately, Problem (\ref{eq:Problem-MM}) is approximated
as 
\begin{align}
\mathop{\max}\limits _{\mathbf{F},\mathbf{e}} & \;\;\sum_{g=1}^{G}f_{g}\left(\mathbf{F},\mathbf{e}\right)\nonumber \\
{\rm s.t.} & \thinspace\thinspace\thinspace\thinspace\mathbf{F}\in\mathcal{S}_{F},\mathbf{e}\in\mathcal{S}_{e}.\label{eq:Problem-2}
\end{align}

This problem is still a biconvex problem of $\mathbf{F}$ and $\mathbf{e}$,
which enables us to alternately update $\mathbf{F}$ and $\mathbf{e}$
by adopting the alternating optimization method.

\subsection{Optimizing the Precoding Matrix $\mathbf{F}$}

Given $\mathbf{e}$, the subproblem of Problem (\ref{eq:Problem-2})
for the optimization of $\mathbf{F}$ is 
\begin{align}
\mathop{\max}\limits _{\mathbf{F}} & \;\;\sum_{g=1}^{G}f_{g}\left(\mathbf{F}\right)\nonumber \\
{\rm s.t.} & \thinspace\thinspace\thinspace\thinspace\mathbf{F}\in\mathcal{S}_{F}.\label{eq:Problem-f}
\end{align}
Even $f_{g}\left(\mathbf{F}\right)$ is a concave and continuous function
of precoding matrix $\mathbf{F}$, it is still very complex and difficult
to be optimized directly. In this subsection, the surrogate function
of $f_{g}\left(\mathbf{F}\right)$ in the MM algorithm framework is
given in the following theorem.

\begin{theorem}\label{Theorem-1} Since $f_{g}\left(\mathbf{F}\right)$
is twice differentiable and concave, we minorize $f_{g}\left(\mathbf{F}\right)$
at any fixed $\mathbf{F}^{n}$ with a quadratic function $\widetilde{f}_{g}(\mathbf{F}|\mathbf{F}^{n})$
satisfying conditions (A1)-(A4), as follows 
\begin{align}
\widetilde{f}_{g}(\mathbf{F}|\mathbf{F}^{n})=2\textrm{\ensuremath{\mathrm{Re}}}\left\{ \mathrm{Tr}\left[\mathbf{U}_{g}^{\mathrm{H}}\mathbf{F}\right]\right\} +\alpha_{g}\mathrm{Tr}\left[\mathbf{F}^{\mathrm{H}}\mathbf{F}\right]+\textrm{consF}_{g},\label{quadratic-f}
\end{align}
where 
\begin{align}
 & \mathbf{U}_{g}=\sum_{k\in\mathcal{K}_{g}}g_{k}(\mathbf{F}^{n})(\mathbf{C}_{k}-\mathbf{B}_{k}^{\mathrm{H}}\mathbf{F}^{n})-\alpha_{g}\mathbf{F}^{n},\label{eq:U}\\
 & g_{k}(\mathbf{F}^{n})=\frac{\mathrm{exp}\left\{ -\mu_{g}\widetilde{R}_{k}\left(\mathbf{F}^{n}\right)\right\} }{\sum_{k\in\mathcal{K}_{g}}\mathrm{exp}\left\{ -\mu_{g}\widetilde{R}_{k}\left(\mathbf{F}^{n}\right)\right\} },k\in\mathcal{K}_{g},\label{g_k_f}\\
 & \alpha_{g}=-\underset{k\in\mathcal{K}_{g}}{\mathrm{max}}\left\{ b_{k}\mathbf{e}^{\mathrm{H}}\mathbf{H}_{k}\mathbf{H}_{k}^{\mathrm{H}}\mathbf{e}\right\} -2\mu_{g}\underset{k\in\mathcal{K}_{g}}{\mathrm{max}}\left\{ tp_{k}\right\} ,\label{alpha}\\
 & tp_{k}=P_{\mathrm{T}}b_{k}^{2}|\mathbf{e}^{\mathrm{H}}\mathbf{H}_{k}\mathbf{H}_{k}^{\mathrm{H}}\mathbf{e}|^{2}+||\mathbf{C}_{k}||_{F}^{2}+2\sqrt{P_{\mathrm{T}}}||\mathbf{B}_{k}\mathbf{C}_{k}||_{F},\label{eq:tpk}\\
 & \textrm{consF}_{g}=f_{g}(\mathbf{F}^{n})+\alpha_{g}\mathrm{Tr}\left[(\mathbf{F}^{n})^{\mathrm{H}}\mathbf{F}^{n}\right]-2\textrm{\ensuremath{\mathrm{Re}}}\left\{ \mathrm{Tr}\left[\mathbf{D}_{g}^{\mathrm{H}}\mathbf{F}^{n}\right]\right\} .\label{constant-f}
\end{align}

\end{theorem}

\textbf{\textit{Proof: }}Please refer to Appendix \ref{subsec:The-proof-of-2}.\hspace{3cm}$\blacksquare$

Upon replacing the objective function of Problem (\ref{eq:Problem-f})
with (\ref{quadratic-f}), we obtain the following surrogate problem
\begin{align}
\mathop{\max}\limits _{\mathbf{F}} & \;\;\sum_{g=1}^{G}\left(2\textrm{\ensuremath{\mathrm{Re}}}\left\{ \mathrm{Tr}\left[\mathbf{U}_{g}^{\mathrm{H}}\mathbf{F}\right]\right\} +\alpha_{g}\mathrm{Tr}\left[\mathbf{F}^{\mathrm{H}}\mathbf{F}\right]+\textrm{consF}_{g}\right)\nonumber \\
{\rm s.t.} & \thinspace\thinspace\thinspace\thinspace\mathbf{F}\in\mathcal{S}_{F}.\label{eq:Problem-f-final}
\end{align}

The optimal $\mathbf{F}^{n+1}$ could be obtained by introducing a
Lagrange multiplier $\tau\geq0$ associated with the power constraint,
yielding the Lagrange function 
\begin{align}
\mathcal{L}\text{(\ensuremath{\mathbf{F}},\ensuremath{\tau})} & =2\mathrm{Re}\left\{ \mathrm{Tr}\left[\sum_{g=1}^{G}\mathbf{U}_{g}^{\mathrm{H}}\mathbf{F}\right]\right\} +\sum_{g=1}^{G}\alpha_{g}\mathrm{Tr}\left[\mathbf{F}^{\mathrm{H}}\mathbf{F}\right]\nonumber \\
 & +\sum_{g=1}^{G}\textrm{consF}_{g}-\tau\left(\mathrm{Tr}\left[\mathbf{F}^{\mathrm{H}}\mathbf{F}\right]-P_{\mathrm{T}}\right).\label{eq:L-mm-F}
\end{align}

By setting the first-order derivative of $\mathcal{L}\text{(\ensuremath{\mathbf{F}},\ensuremath{\tau})}$
w.r.t. $\mathbf{F}^{*}$ to zero, we have 
\[
\frac{\partial\mathcal{L}\text{(\ensuremath{\mathbf{F}})}}{\partial\mathbf{F}^{*}}=\mathbf{0}.
\]
Then the globally optimal solution of $\mathbf{F}$ in iteration $n$
can be derived as 
\begin{equation}
\mathbf{F}^{n+1}=\frac{1}{\tau-\sum_{g=1}^{G}\alpha_{g}}\sum_{g=1}^{G}\mathbf{U}_{g}.\label{eq:optimal-F-tao}
\end{equation}

By substituting (\ref{eq:optimal-F-tao}) into the power constraint,
one has 
\begin{equation}
\frac{\mathrm{Tr}\left[\left(\sum_{g=1}^{G}\mathbf{U}_{g}\right)^{\mathrm{H}}\left(\sum_{g=1}^{G}\mathbf{U}_{g}\right)\right]}{(\tau-\sum_{g=1}^{G}\alpha_{g})^{2}}\leq P_{\mathrm{T}}.\label{eq:constraint-tao}
\end{equation}
It is obvious that the left hand side of (\ref{eq:constraint-tao})
is a decreasing function of $\tau$. 
\begin{itemize}
\item If the power constraint inequality (\ref{eq:constraint-tao}) holds
when $\tau=0$, then 
\begin{equation}
\mathbf{F}^{n+1}=\frac{-1}{\sum_{g=1}^{G}\alpha_{g}}\sum_{g=1}^{G}\mathbf{U}_{g}.
\end{equation}
\item Otherwise, there must exist a $\tau>0$ that (\ref{eq:constraint-tao})
holds with equality, then 
\end{itemize}
\begin{equation}
\mathbf{F}^{n+1}=\sqrt{\frac{P_{\mathrm{T}}}{\mathrm{Tr}\left[\left(\sum_{g=1}^{G}\mathbf{U}_{g}\right)^{\mathrm{H}}\left(\sum_{g=1}^{G}\mathbf{U}_{g}\right)\right]}}\sum_{g=1}^{G}\mathbf{U}_{g}.
\end{equation}

\subsection{Optimizing the Reflection Coefficient Vector $\mathbf{e}$}

Given $\mathbf{F}$, the subproblem of Problem (\ref{eq:Problem-2})
for the optimization of $\mathbf{e}$ is 
\begin{align}
\mathop{\max}\limits _{\mathbf{e}} & \;\;\sum_{g=1}^{G}f_{g}\left(\mathbf{e}\right)\nonumber \\
{\rm s.t.} & \thinspace\thinspace\thinspace\thinspace\mathbf{e}\in\mathcal{S}_{e}.\label{eq:Problem-e}
\end{align}
Upon adopting the MM algorithm framework, we first need to find a
minorizing function of $f_{g}\left(\mathbf{e}\right)$ and denote
it as $\widehat{f}_{g}(\mathbf{e}|\mathbf{e}^{n})$. Since $\mathcal{S}_{e}$
is a non-convex set, we should modify (A3) so as to claim stationarity
convergence \cite{pang2007partially,pang2017computing}: 
\[
\widehat{f}_{g}^{'}(\mathbf{e}|\mathbf{e}^{n};\mathbf{d})|_{\mathbf{e}=\mathbf{e}^{n}}=f_{g}^{'}(\mathbf{e}^{n};\mathbf{d}),\forall\mathbf{d}\in\mathcal{T}_{\mathcal{S}_{e}}(\mathbf{e}^{n}),
\]
where $\mathcal{T}_{\mathcal{S}_{e}}(\mathbf{e}^{n})$ is the Boulingand
tangent cone of $\mathcal{S}_{e}$ at $\mathbf{e}^{n}$. Therefore
$\widehat{f}_{g}(\mathbf{e}|\mathbf{e}^{n})$ is given in the following
theorem.

\begin{theorem}\label{Theorem-2} Since $f_{g}\left(\mathbf{e}\right)$
is twice differentiable and concave, we minorize $f_{g}\left(\mathbf{e}\right)$
at any fixed $\mathbf{e}^{n}$ with a function $\widehat{f}_{g}(\mathbf{e}|\mathbf{e}^{n})$
satisfying conditions (A1)-(A4), as follows 
\begin{align}
\widetilde{f}_{g}(\mathbf{e}|\mathbf{e}^{n})=2\textrm{\ensuremath{\mathrm{Re}}}\left\{ \mathbf{u}_{g}^{\mathrm{H}}\mathbf{e}\right\} +\textrm{consE}_{g},\label{quadratic-e}
\end{align}
where 
\begin{align}
 & \mathbf{u}_{g}=\sum_{k\in\mathcal{K}_{g}}g_{k}(\mathbf{e}^{n})(\mathbf{a}_{k}-\mathbf{A}_{k}^{\mathrm{H}}\mathbf{e}^{n})-\beta_{g}\mathbf{e}^{n},\label{u}\\
 & g_{k}(\mathbf{e}^{n})=\frac{\mathrm{exp}\left\{ -\mu_{g}\widetilde{R}_{k}\left(\mathbf{e}^{n}\right)\right\} }{\sum_{k\in\mathcal{K}_{g}}\mathrm{exp}\left\{ -\mu_{g}\widetilde{R}_{k}\left(\mathbf{e}^{n}\right)\right\} },k\in\mathcal{K}_{g},\label{eq:gke}\\
 & \beta_{g}=-\mathrm{max}_{k\in\mathcal{K}_{g}}\left\{ \lambda_{\mathrm{max}}(\mathbf{A}_{k})\right\} -2\mu_{g}\mathrm{max}_{k\in\mathcal{K}_{g}}\left\{ tp2_{k}\right\} ,\label{beta}\\
 & tp2_{k}=||\mathbf{a}_{k}||_{2}^{2}+(M+1)\lambda_{\mathrm{max}}(\mathbf{A}_{k}\mathbf{A}_{k}^{\mathrm{H}})+2||\mathbf{A}_{k}\mathbf{a}_{k}||_{1},\label{eq:tp2k}\\
 & \textrm{consE}_{g}=f_{g}(\mathbf{e}^{n})+2(M+1)\beta_{g}-2\textrm{\ensuremath{\mathrm{Re}}}\left\{ \mathbf{d}_{g}^{\mathrm{H}}\mathbf{e}^{n}\right\} .\label{eq:constant-e}
\end{align}

\end{theorem}

\textbf{\textit{Proof: }}Please refer to Appendix \ref{subsec:The-proof-of-4}.\hspace{3cm}$\blacksquare$

Upon replacing the objective function of Problem (\ref{eq:Problem-e})
by (\ref{quadratic-e}), we obtain the following surrogate problem
as 
\begin{align}
\mathop{\max}\limits _{\mathbf{e}} & \;\;\sum_{g=1}^{G}\left(2\textrm{\ensuremath{\mathrm{Re}}}\left\{ \mathbf{u}_{g}^{\mathrm{H}}\mathbf{e}\right\} +\textrm{consE}_{g}\right)\nonumber \\
{\rm s.t.} & \thinspace\thinspace\thinspace\thinspace\mathbf{e}\in\mathcal{S}_{e}.\label{eq:Problem-e-final}
\end{align}
Then, the globally optimal solution of $\mathbf{e}$ at the $n^{\mathrm{th}}$
iteration is 
\begin{equation}
\mathbf{e}^{n+1}=\exp\left\{ j\angle\left(\left(\sum_{g=1}^{G}\mathbf{u}_{g}\right)/\left[\sum_{g=1}^{G}\mathbf{u}_{g}\right]_{M+1}\right)\right\} ,\label{eq:MM-e}
\end{equation}
where $\exp\left\{ j\angle\left(\cdot\right)\right\} $ is an element-wise
operation.

\subsection{Low-complexity algorithm design}

In this section, we adopt alternating optimization algorithm to alternately
optimize precoding matrix $\mathbf{F}$ and reflection coefficient
vector $\mathbf{e}$. Note that the tightness of the lower bounds
$\alpha_{g}$ in (\ref{alpha}) and $\beta_{g}$ in (\ref{beta})
affects the performance of the convergence speed. Here, we adopt SQUAREM
\cite{varadhan2008SQUAREM} to accelerate the convergence speed of
our proposed algorithm, which is summarized in Algorithm \ref{Algorithm-MM}.

Let $\mathcal{M}_{F}(\cdot)$ denote the nonlinear fixed-point iteration
map of the MM algorithm of $\mathbf{F}$ in (\ref{eq:optimal-F-tao}),
i.e., $\mathbf{F}^{n+1}=\mathcal{M}_{F}(\mathbf{F}^{n})$, and $\mathcal{M}_{e}(\cdot)$
of $\mathbf{e}$ in (\ref{eq:MM-e}), i.e., $\mathbf{e}^{n+1}=\mathcal{M}_{e}(\mathbf{e}^{n})$.
$\mathcal{\mathcal{P_{S}}}(\cdot)$ is project operation to force
wayward points to satisfy their nonlinear constraints. For the power
constraint in Problem (\ref{eq:Problem-f-final}), the projection
can be done by using the function $\frac{\left(\cdot\right)}{||\cdot||_{F}}||\mathbf{F}_{2}||_{F}$
to the solution matrix, e.g., $\mathcal{\mathcal{P_{S}}}(\mathbf{X})=\frac{\left(\mathbf{X}\right)}{||\mathbf{X}||_{F}}||\mathbf{F}_{2}||_{F}$.
For the unit-modulus constraints in Problem (\ref{eq:Problem-e-final}),
it can be obtained by using function $\exp\left\{ j\angle(\cdot)\right\} $
element-wise to the solution vector. Steps 10 to 13 and steps 21 to
24 are to maintain the ascent property of the proposed algorithm.

\begin{algorithm}
\caption{Low-complexity MM algorithm}
\label{Algorithm-MM} \begin{algorithmic}[1] \REQUIRE Initialize
$\mathbf{F}^{0}$ and $\mathbf{e}^{0}$, and $n=0$.

\REPEAT

\STATE Set $\mathbf{e}=\mathbf{e}^{n}$;

\STATE $\mathbf{F}_{1}=\mathcal{M}_{F}(\mathbf{F}^{n})$;

\STATE $\mathbf{F}_{2}=\mathcal{M}_{F}(\mathbf{F}_{1})$;

\STATE $\mathbf{J}_{1}=\mathbf{F}_{1}-\mathbf{F}^{n}$;

\STATE $\mathbf{J}_{2}=\mathbf{F}_{2}-\mathbf{F}_{1}-\mathbf{\mathbf{J}}_{1}$;

\STATE $\omega=-\frac{||\mathbf{J}_{1}||_{F}}{||\mathbf{J}_{2}||_{F}}$;

\STATE $\mathbf{F}^{n+1}=-\mathcal{\mathcal{P_{S}}}(\mathbf{F}^{n}-2\omega\mathbf{J}_{1}+\omega^{2}\mathbf{J}_{2})$;

\WHILE {$F\left(\mathbf{F}^{n+1}\right)<F\left(\mathbf{F}^{n}\right)$}

\STATE $\thinspace\thinspace\thinspace\thinspace\thinspace\thinspace\thinspace\omega=(\omega-1)/2$;

\STATE $\thinspace\thinspace\thinspace\thinspace\thinspace\thinspace\mathbf{F}^{n+1}=-\mathcal{\mathcal{P_{S}}}(\mathbf{F}^{n}-2\omega\mathbf{J}_{1}+\omega^{2}\mathbf{J}_{2})$;

\ENDWHILE

\STATE Set $\mathbf{F}=\mathbf{F}^{n+1}$;

\STATE $\mathbf{e}_{1}=\mathcal{M}_{e}(\mathbf{e}^{n})$;

\STATE $\mathbf{e}_{2}=\mathcal{M}_{e}(\mathbf{e}_{1})$;

\STATE $\mathbf{j}_{1}=\mathbf{e}_{1}-\mathbf{e}^{n}$;

\STATE $\mathbf{j}_{2}=\mathbf{e}_{2}-\mathbf{e}_{1}-\mathbf{\mathbf{j}}_{1}$;

\STATE $\omega=-\frac{||\mathbf{j}_{1}||_{F}}{||\mathbf{j}_{2}||_{F}}$;

\STATE $\mathbf{e}^{n+1}=-\mathcal{\mathcal{P_{S}}}(\mathbf{e}^{n}-2\omega\mathbf{j}_{1}+\omega^{2}\mathbf{j}_{2})$;

\WHILE {$F\left(\mathbf{e}^{n+1}\right)<F\left(\mathbf{e}^{n}\right)$}

\STATE $\thinspace\thinspace\thinspace\thinspace\thinspace\thinspace\thinspace\omega=(\omega-1)/2$;

\STATE $\thinspace\thinspace\thinspace\thinspace\thinspace\thinspace\mathbf{e}^{n+1}=-\mathcal{\mathcal{P_{S}}}(\mathbf{e}^{n}-2\omega\mathbf{j}_{1}+\omega^{2}\mathbf{j}_{2})$;

\ENDWHILE

\STATE $n\leftarrow n+1$;

\UNTIL The value of function $F\left(\mathbf{F},\mathbf{e}\right)$
in (\ref{eq:Problem-original}) converges. \end{algorithmic} 
\end{algorithm}

\subsection{Complexity Analysis}

The computational complexity of Algorithm \ref{Algorithm-MM} is composed
of the nonlinear fixed-point iteration maps $\mathcal{M}_{F}(\cdot)$
and $\mathcal{M}_{e}(\cdot)$. In $\mathcal{M}_{F}(\cdot)$, the computational
complexity of $\mathbf{U}_{g}$ in (\ref{eq:tpk}) mainly comes from
$g_{k}(\mathbf{F}^{n})$ in (\ref{g_k_f}) and $\alpha_{g}$ in (\ref{alpha}).
Firstly, the computational complexity of $g_{k}(\mathbf{F}^{n})$
is of order $\mathcal{O}(|\mathcal{K}_{g}|(2MNG+3NG))$ since there
are $|\mathcal{K}_{g}|$ $\widetilde{R}_{k}\left(\mathbf{F}^{n}\right)$
in (\ref{eq:Rate-surrogate}) of order $\mathcal{O}(2MNG+3NG)$. Then
each $tp_{k}$ in (\ref{eq:tpk}) is of complexity $\mathcal{O}(4N^{3}+2N^{2}K-NK+4MN)$
neglecting the lower-order terms, thus $\alpha_{g}$ is of order $\mathcal{O}(|\mathcal{K}_{g}|(4N^{3}+2N^{2}K+4MN))$.
Therefore, the approximate complexity of $\mathcal{M}_{F}(\cdot)$
is $\mathcal{O}(4N^{3}K+2N^{2}K^{2}+2MNGK)$ neglecting the lower-order
terms. In $\mathcal{M}_{e}(\cdot)$, the computational complexity
of $g_{k}(\mathbf{e}^{n})$ in (\ref{eq:gke}) is the same as $g_{k}(\mathbf{F}^{n})$,
which is of complexity $\mathcal{O}(|\mathcal{K}_{g}|(2MNG+3NG))$.
Furthermore, the eigenvalue operations $\lambda_{\mathrm{max}}(\mathbf{A}_{k})$
and $\lambda_{\mathrm{max}}(\mathbf{A}_{k}\mathbf{A}_{k}^{\mathrm{H}})$
of order $\mathcal{O}((M+1)^{3})$ contribute to the main complexity
of calculating $\beta_{g}$ in (\ref{beta}), which is of order $\mathcal{O}(|\mathcal{K}_{g}|(M+1)^{3})$.
Neglecting the lower-order terms, the approximate complexity of $\mathcal{M}_{e}(\cdot)$
is $\mathcal{O}(2MNGK+K(M+1)^{3})$. Eventually, the approximate complexity
of Algorithm \ref{Algorithm-MM} per iteration is $\mathcal{O}(4N^{3}K+2N^{2}K^{2}+3MNGK+K(M+1)^{3})$,
neglecting the lower-order terms.

The computational complexity of the proposed two algorithms are summarized
and compared in Table I. Comparing with Algorithm \ref{Algorithm-SOCP}
based on SOCP, Algorithm \ref{Algorithm-MM} has a lower computational
complexity and requires less CPU time, which will be shown in the
following section. 
\begin{table*}
\centering \centering \caption{Complexity analysis of the proposed MM algorithms}
\begin{tabular}{|c|c|c|}
\hline 
\textbf{Algorithm}  & SOCP-based MM algorithm  & Low-complexity MM algorithm \tabularnewline
\hline 
\textbf{Complexity}  & $\mathcal{O}(N^{3}K^{3}+NK^{4.5}+N^{3}K^{5.5}+MK^{3.5}+M^{3}K^{2.5})$  & $\mathcal{O}(4N^{3}K+2N^{2}K^{2}+3MNGK+K(M+1)^{3})$ \tabularnewline
\hline 
\end{tabular}
\end{table*}

\subsection{Convergence Analysis}

In each iteration, we adopt the MM algorithm to update each set of
variables. The monotonicity of the MM algorithm has been proved in
\cite{MM} and \cite{jacobson2007MM}. In the following, we claim
the monotonicity of Algorithm \ref{Algorithm-MM}. At the $n^{\mathrm{th}}$
iteration, with given $\mathbf{e}^{n}$, we have 
\[
f_{g}(\mathbf{F}^{n},\mathbf{e}^{n})=\widetilde{f}_{g}(\mathbf{F}^{n},\mathbf{F}^{n})\leq\widetilde{f}_{g}(\mathbf{F}^{n+1},\mathbf{F}^{n})\leq f_{g}(\mathbf{F}^{n+1},\mathbf{e}^{n}),
\]
where the first equality follows from (A1), the first inequality follows
from (\ref{eq:Problem-f-final}), and the second one follows from
(A2). Subsequently, with given $\mathbf{F}^{n+1}$, it is straightforward
to have 
\[
f_{g}(\mathbf{F}^{n+1},\mathbf{e}^{n})=\widehat{f}_{g}(\mathbf{e}^{n},\mathbf{e}^{n})\leq\widehat{f}_{g}(\mathbf{e}^{n+1},\mathbf{e}^{n})\leq f_{g}(\mathbf{F}^{n+1},\mathbf{e}^{n+1}).
\]
Therefore, the objective function values $\{f_{g}(\mathbf{F}^{n+1},\mathbf{e}^{n+1})\}$
generated during the procedure of the AO algorithm are monotonically
increasing.

Let $\{\mathbf{F}^{n}\}$ be the sequence generated by the proposed
algorithm. Since $\mathcal{S}_{F}$ is a convex set, every limit point
of $\{\mathbf{F}^{n}\}$ is a d-stationary point of Problem (\ref{eq:Problem-original}),
and the limit point $\mathbf{F}^{\infty}$ satisfies 
\[
f_{g}^{'}(\mathbf{F}^{\infty};\mathbf{d})\leq0,\forall\mathbf{d}\thinspace\thinspace\mathrm{\textrm{with}}\thinspace\thinspace\mathbf{F}^{\infty}+\mathbf{d}\in\mathcal{S}_{F}.
\]
The proof of converging to a d-stationary point can be found in \cite{razaviyayn2013covergence}.

Let $\{\mathbf{e}^{n}\}$ be the sequence generated by the proposed
algorithm. Since $\mathcal{S}_{e}$ is a non-convex set, every limit
point of $\{\mathbf{e}^{n}\}$ is a B-stationary point of Problem
(\ref{eq:Problem-original}), and the limit point $\mathbf{e}^{\infty}$
satisfies 
\[
f_{g}^{'}(\mathbf{e}^{\infty};\mathbf{d})\leq0,\forall\mathbf{d}\in\mathcal{T}_{\mathcal{S}_{e}}(\mathbf{e}^{\infty}).
\]
The proof of converging to a B-stationary point can be found in \cite{pang2007partially}
and \cite{pang2017computing}.

The property of the converged solution of Algorithm \ref{Algorithm-MM}
is shown in the following Theorem.

\begin{theorem}\label{Theorem-MM-kkt} The optimal solution converges
to a KKT point of Problem (\ref{eq:Problem-2}).

\end{theorem}

\textbf{\textit{Proof:}} Please refer to Appendix \ref{subsec:The-proof-of-6}.\hspace{3cm}$\blacksquare$

\section{Simulation results and discussions}

\subsection{Simulation Setup}

\begin{figure}
\centering \includegraphics[width=3in,height=1.3in]{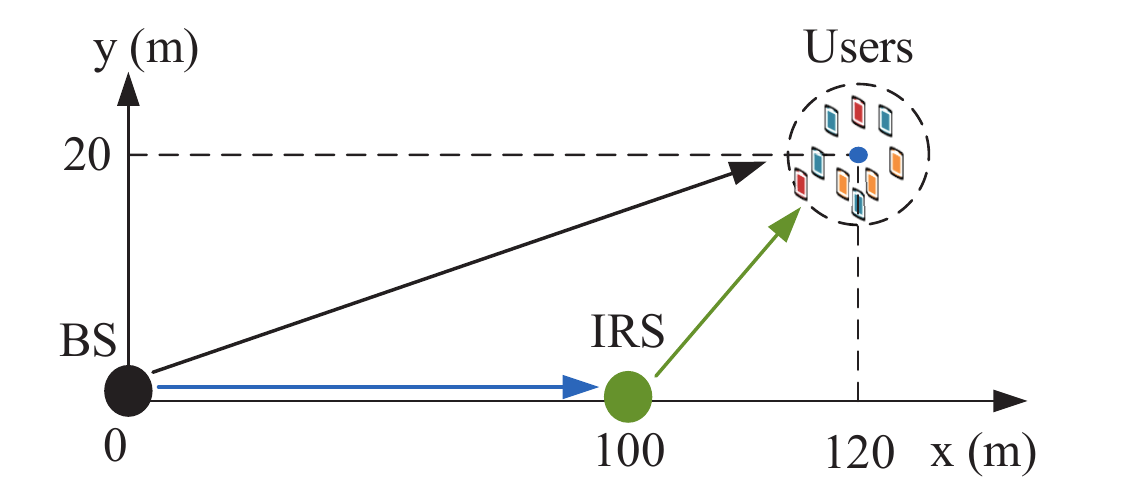}
\caption{The simulated system setup.}
\label{simulated-model} 
\end{figure}

In this section, extensive simulation results are provided to evaluate
the performance of our proposed algorithms for an IRS-aided multigroup
multicast MISO communication system. All experiments are performed
on a PC with a 1.99 GHz i7-8550U CPU and 16 GB RAM. Each point in
the following figures is obtained by averaging over 100 independent
trials. The simulated model in Fig. \ref{simulated-model} is as follows:
The BS locating at (0 m, 0 m) employs a uniform linear array (ULA)
with $N$ antennas and the IRS locating at (100 m, 0 m) is equipped
with a uniform planar array (UPA) with $M$ reflecting elements, where
the width of the UPA is fixed at 4 and the length is $M/4$. All users
are randomly distributed in a circle centered at (120 m, 20 m) with
radius 10 m.

The large-scale path loss is $\mathrm{PL}=-30-10\alpha\log_{10}(d)$
dB, in which $d$ is the link length in meters and the path loss exponents
for the BS-IRS link, the IRS-user link, and the BS-user link are set
as $\alpha_{\mathrm{BI}}=\alpha_{\mathrm{IU}}=2$ and $\alpha_{\mathrm{BU}}=4$,
respectively \cite{emil-pathloss}. The small-scale fading in $[\mathbf{H_{\mathrm{dr}}},\{\mathbf{h}_{\mathrm{d},k}\}_{\forall k\in\mathcal{K}}]$
is assumed to follow Rayleigh distribution with zero-mean and unit
variance due to the fact of the large lengths of the BS-IRS link and
the BS-user link, while the small-scale fading in $\{\mathbf{h}_{\mathrm{r},k}\}_{\forall k\in\mathcal{K}}$
is assumed to be Rican fading with Ricean factor $\kappa_{\mathrm{IU}}=10$.
The line-of-sight (LoS) components are modeled as the product of the
steering vectors of the transceivers and the non-LoS components are
drawn from a Rayleigh distribution. Unless otherwise stated, the other
parameters are set as: Transmission bandwidth of 10 MHz, noise power
density of $-174$ dBm/Hz, convergence accuracy of $\epsilon=10^{-6}$,
smoothing parameter of $\mu_{g}=100$ \cite{xu2001smoothing}, $N=4$,
$N=16$, $G=|\mathcal{K}_{g}|=2$.

We use \textbf{IRS-Alg. 1} to represent Algorithm \ref{Algorithm-SOCP}
and \textbf{IRS-Alg. 2} to represent Algorithm \ref{Algorithm-MM}.
For comparison purposes, we show the performance of the scheme without
IRS, in which the precoding matrix is also obtained by our proposed
two algorithms, denoted as \textbf{NIRS-Alg. 1} and \textbf{NIRS-Alg.
2}, respectively.

\subsection{Baseline Schemes}

Due to the hardware limitation, it is practically difficult to realize
the continuous phase shifts at each reflection element considered
in this work. Hence, two baseline schemes with $2$ bit resolution
are considered in the simulations to investigate the performance loss
of using finite resolution reflection elements. Specifically, with
optimal $\mathbf{e}^{o}$ generated by Algorithm \ref{Algorithm-SOCP}
or Algorithm \ref{Algorithm-MM}, the $m^{th}$ discrete phase shift
can be obtained by 
\[
\theta_{m}^{o}=\arg\min_{\theta\in\mathcal{F}_{\theta}}|\exp\left\{ j\angle\theta\right\} -e_{m}^{o}|,
\]
where $\mathcal{F}_{\theta}=\{0,2\pi/B,...,2\pi(B-1)/B\}$ and $B=2^{2}$.
Therefore, we call the two baseline schemes as \textbf{IRS-Alg. 1,
2 bit }and \textbf{IRS-Alg. 2, 2 bit.}

Besides, IRS is advocated as an energy-efficient device for assisting
wireless communication. Hence, it is necessary to compare the performance
of the IRS-based and the full-duplex amplify-and-forward (AF) relay-based
multigroup multicast systems. To ensure a fair comparison with our
proposed IRS-aided system, the\textbf{ Relay} benchmark scheme, in
which the relay is located at the same place of the IRS, has considered
the same users' locations and channel realizations. Then, the sum
rate maximization problem for the joint design of the precoder $\mathbf{F}$
and the relay beamforming $\mathbf{W}$ is given by 
\begin{align}
\mathop{\max}\limits _{\mathbf{F},\mathbf{W}} & \;\;\sum_{g=1}^{G}\min_{k\in\mathcal{K}_{g}}R_{k}^{relay}\nonumber \\
{\rm s.t.} & \;\;||\mathbf{F}||_{F}^{2}\leq P_{\mathrm{T}}\nonumber \\
 & \;\;||\mathbf{W}\mathbf{H_{\mathrm{dr}}}\mathbf{F}||_{F}^{2}+||\mathbf{W}||_{F}^{2}\sigma_{r}^{2}\leq P_{\mathrm{relay}},\label{eq:relay-problem}
\end{align}
where $R_{k}^{relay}$ is given by 
\[
\log_{2}\left(1+\frac{|(\mathbf{h}_{\mathrm{d},k}^{\mathrm{H}}+\mathbf{h}_{\mathrm{r},k}^{\mathrm{H}}\mathbf{W}\mathbf{H_{\mathrm{dr}}}){\bf f}_{g}|^{2}}{\sum_{i\neq g}^{G}|(\mathbf{h}_{\mathrm{d},k}^{\mathrm{H}}+\mathbf{h}_{\mathrm{r},k}^{\mathrm{H}}\mathbf{W}\mathbf{H_{\mathrm{dr}}}){\bf f}_{i}|^{2}+||\mathbf{h}_{\mathrm{r},k}^{\mathrm{H}}\mathbf{W}||_{2}^{2}\sigma_{r}^{2}+\sigma_{k}^{2}}\right).
\]
Here, $P_{\mathrm{relay}}$ is the maximum available transmit power
at the relay, $\sigma_{r}^{2}$ is the noise power received by the
relay, and the digital relay beamforming $\mathbf{W}$ is assumed
to be a diagonal matrix.

The AO method is adopted to solve the above problem. Basically, we
extend the SCA method in \cite{Tervo2018tradeoff} to alternately
update each variable in Problem (\ref{eq:relay-problem}).

\subsection{Convergence of the Proposed Algorithms}

Consider the fact of the nonconvexity of Problem (\ref{eq:Problem-original}),
different initial points may result in different locally optimal solutions
obtained by the our proposed algorithms. By testing 30 randomly channel
realizations, Fig. 3 illustrates the impact of the initializations
on the performance of the proposed algorithms. The initializations
of IRS-Alg. 1 and IRS-Alg. 2 are: ${\bf F}$ is initialized by uniformly
allocating maximum transmit power, ${\bf e}$ is initialized by setting
each entry to 1. IRS-Alg. 1-EXH (IRS-Alg. 2-EXH) refers to the best
initial point of 1000 random initial points for each channel realization.
It can be seen that the sum rate of IRS-Alg. 1 (IRS-Alg. 2) is almost
the same as that of IRS-Alg. 1-EXH (IRS-Alg. 2-EXH), implying that
the simple uniform power allocation of ${\bf F}$ and all-one ${\bf e}$
is a good option for the initialization.

\begin{figure}
\centering \includegraphics[width=3.4in,height=2.6in]{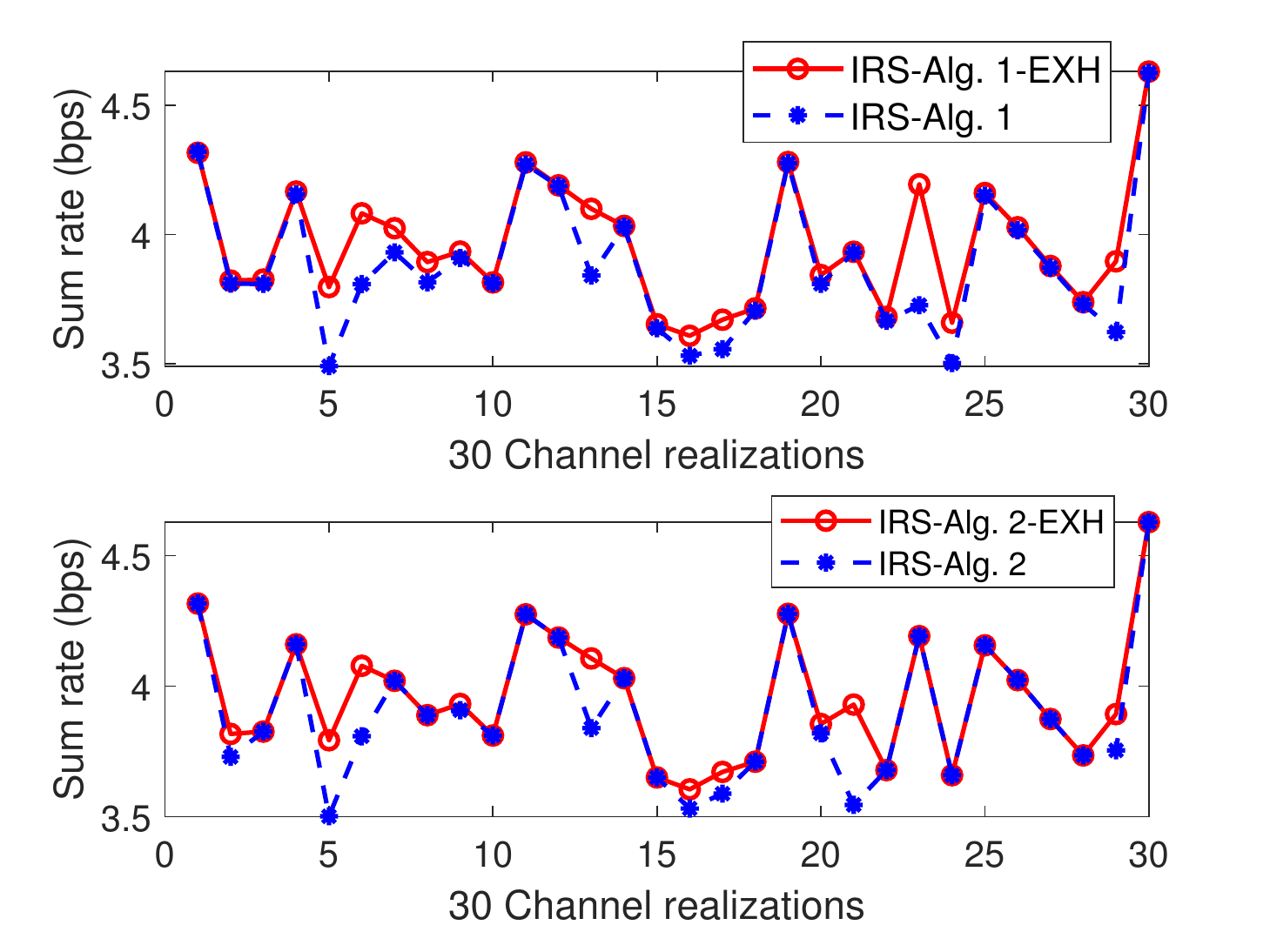}
\caption{The performance comparison of different initialization, when $N=4$,
$N=16$, $G=|\mathcal{K}_{g}|=2$ and $P_{\mathrm{T}}=15$ dBm.}
\label{initialization} 
\end{figure}

\begin{figure}
\centering \subfigure[The sum rate versus iteration number]{ %
\begin{minipage}[t]{0.495\linewidth}%
\centering \includegraphics[width=1.8in]{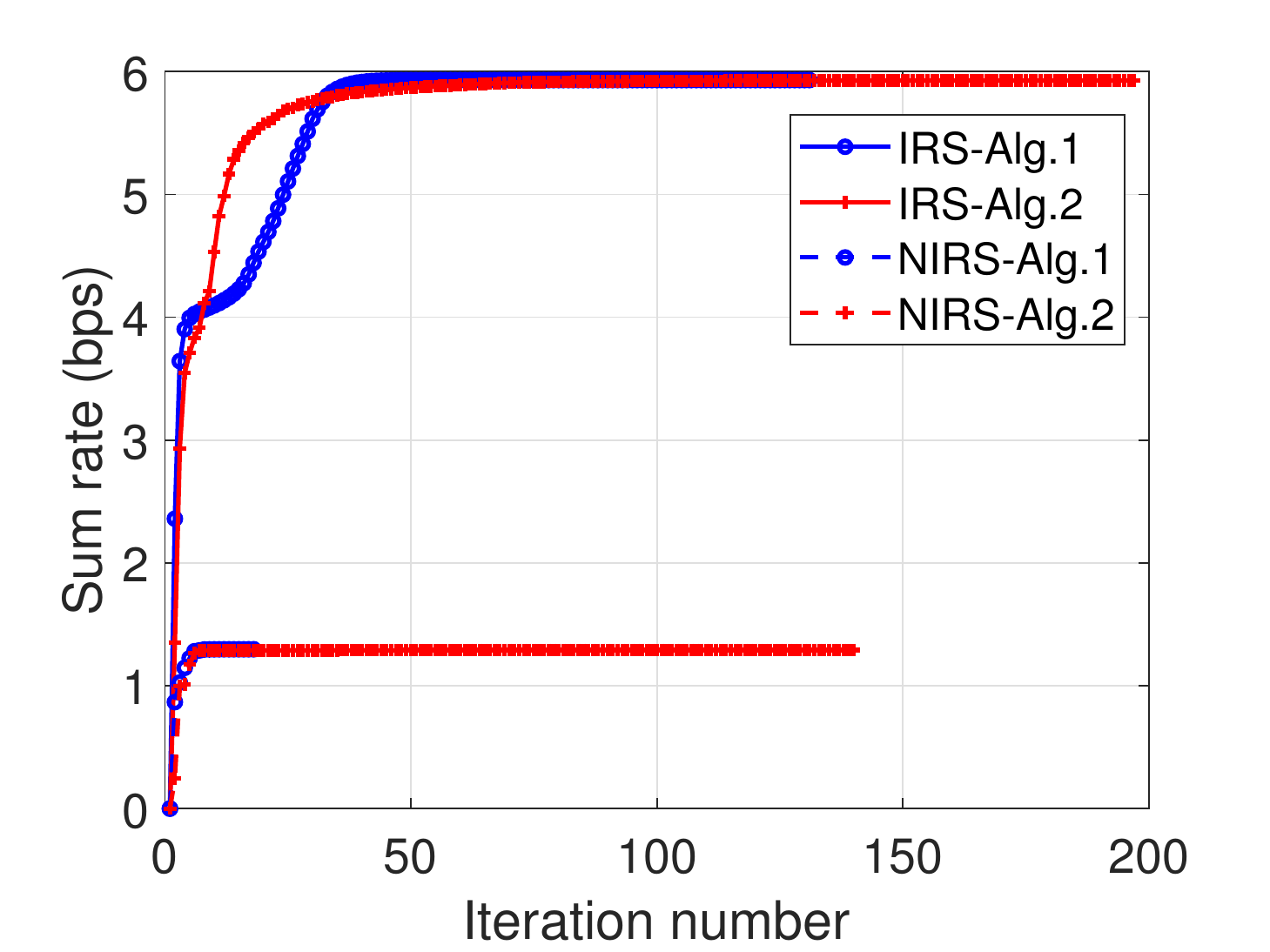} 
\end{minipage}}\subfigure[The sum rate versus CPU time]{ %
\begin{minipage}[t]{0.495\linewidth}%
\centering \includegraphics[width=1.8in]{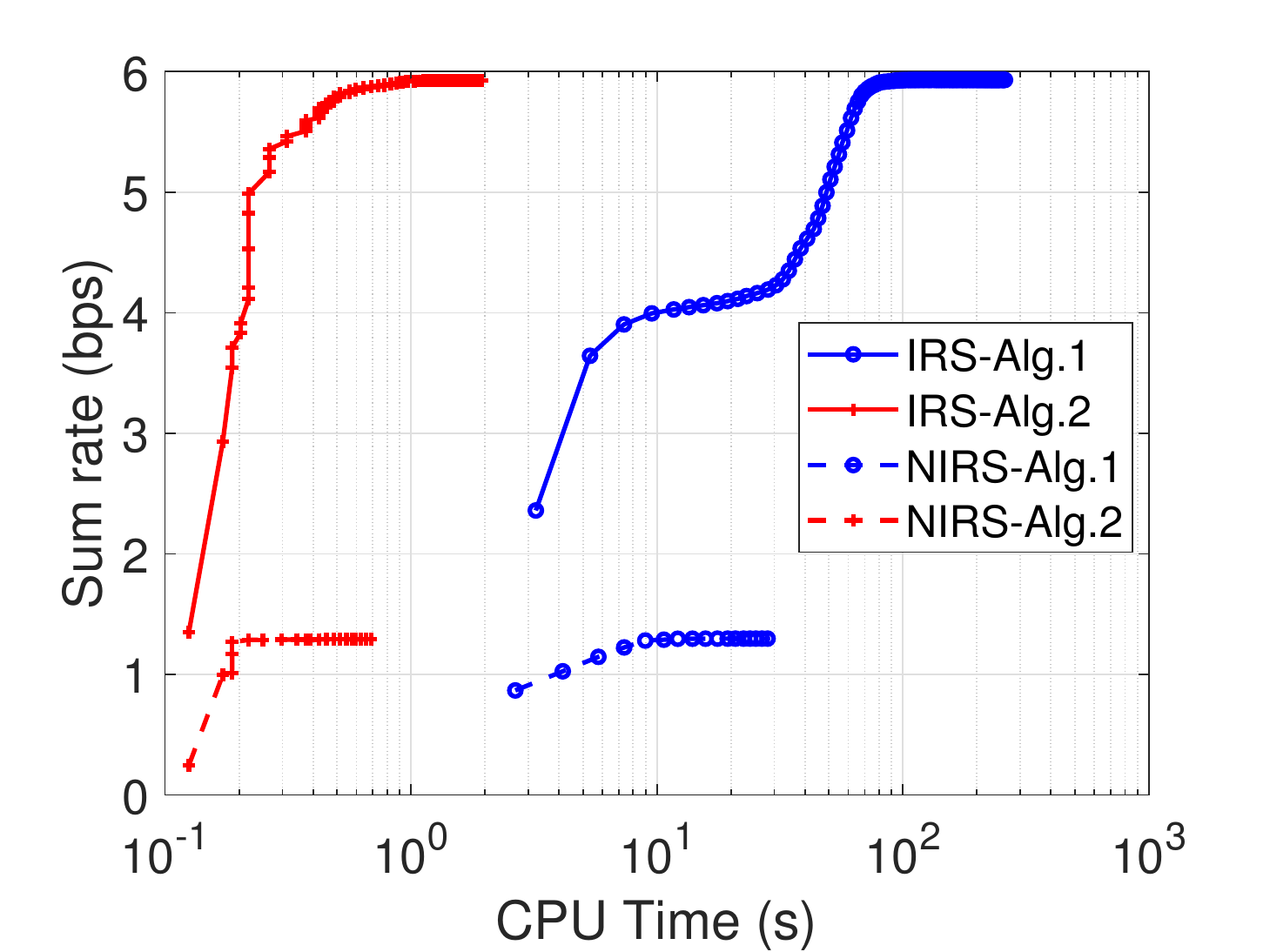} 
\end{minipage}}

\caption{The convergence behaviour of different algorithms, when $N=4$, $N=16$,
$G=|\mathcal{K}_{g}|=2$ and $P_{\mathrm{T}}=20$ dBm.}
\label{convergence} 
\end{figure}

In Fig. \ref{convergence} investigates the convergence behaviour
of various algorithms in terms of the iteration number and the CPU
time when $P_{\mathrm{T}}=20$ dBm. Fig. \ref{convergence}(a) compares
convergence speed in terms of the number of iterations. Only a small
number of iterations are sufficient for Algorithm \ref{Algorithm-SOCP}
to converge for both IRS and NIRS schemes. The reason is that the
lower bound of the original objective function in (\ref{eq:Rate-surrogate})
used in Algorithm \ref{Algorithm-SOCP} is tighter than those in (\ref{quadratic-f})
and (\ref{quadratic-e}) used in Algorithm \ref{Algorithm-MM}. Although
Algorithm \ref{Algorithm-MM} needs more iterations to converge, it
has a fast convergence speed in terms of CPU time shown in Fig. \ref{convergence}(b).
This is because in each iteration of Algorithm \ref{Algorithm-MM},
there always exists closed-form solutions when designing precoding
matrix and reflection coefficient vector. In addition, the optimal
objective function values generated by both algorithms for IRS case
and NIRS case are the same. Therefore, Algorithm \ref{Algorithm-MM}
outperforms Algorithm \ref{Algorithm-SOCP} due to the fact that the
former can generate the same gain with the latter while costing much
less CPU running time

\subsection{IRS vs AF relay Performance Comparison}

\begin{figure}
\centering \subfigure[Sum rate versus transmit power]{\includegraphics[width=3.4in,height=2.6in]{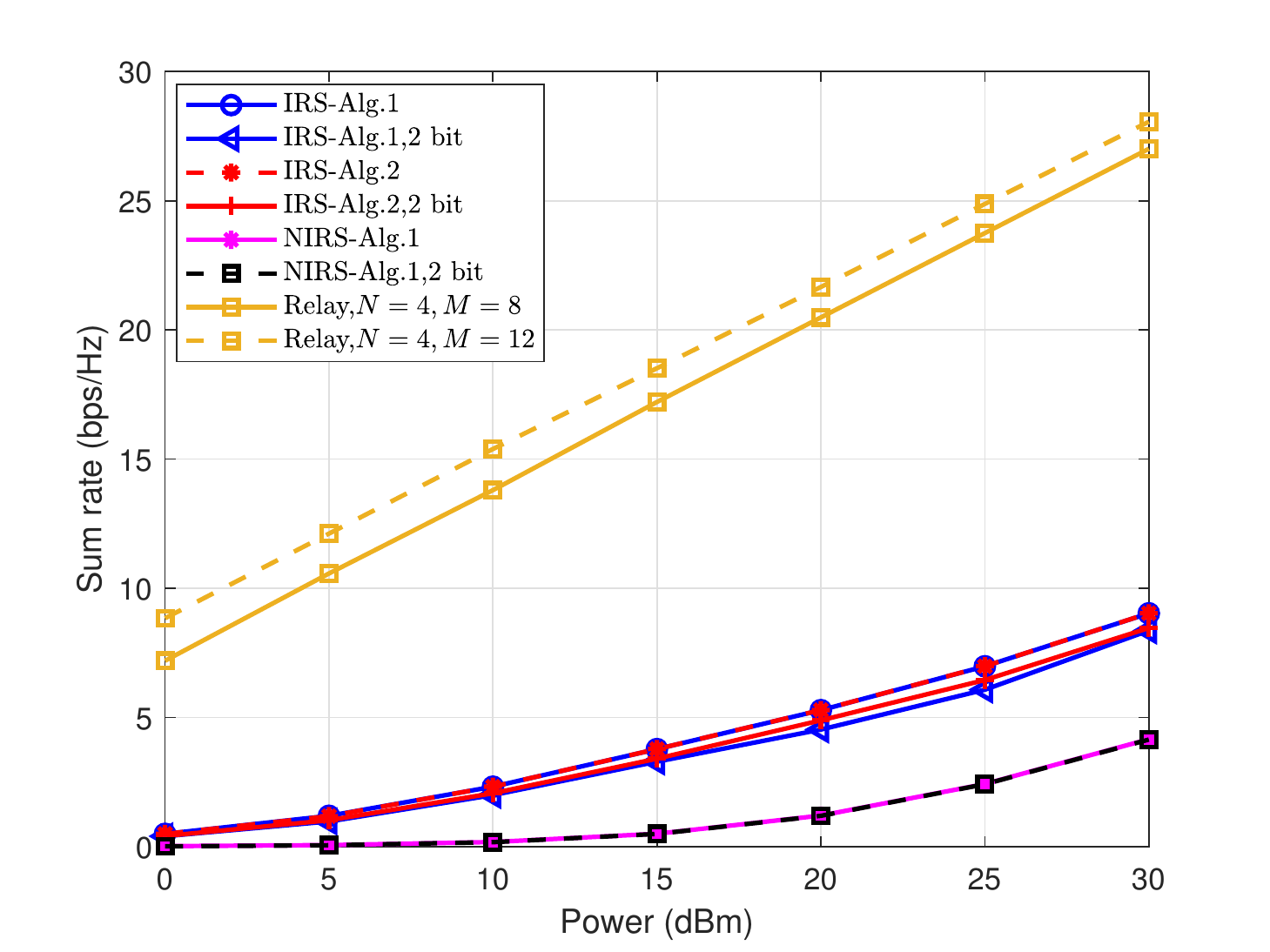}}

\centering \subfigure[Energy efficiency versus transmit power ]{\includegraphics[width=3.4in,height=2.6in]{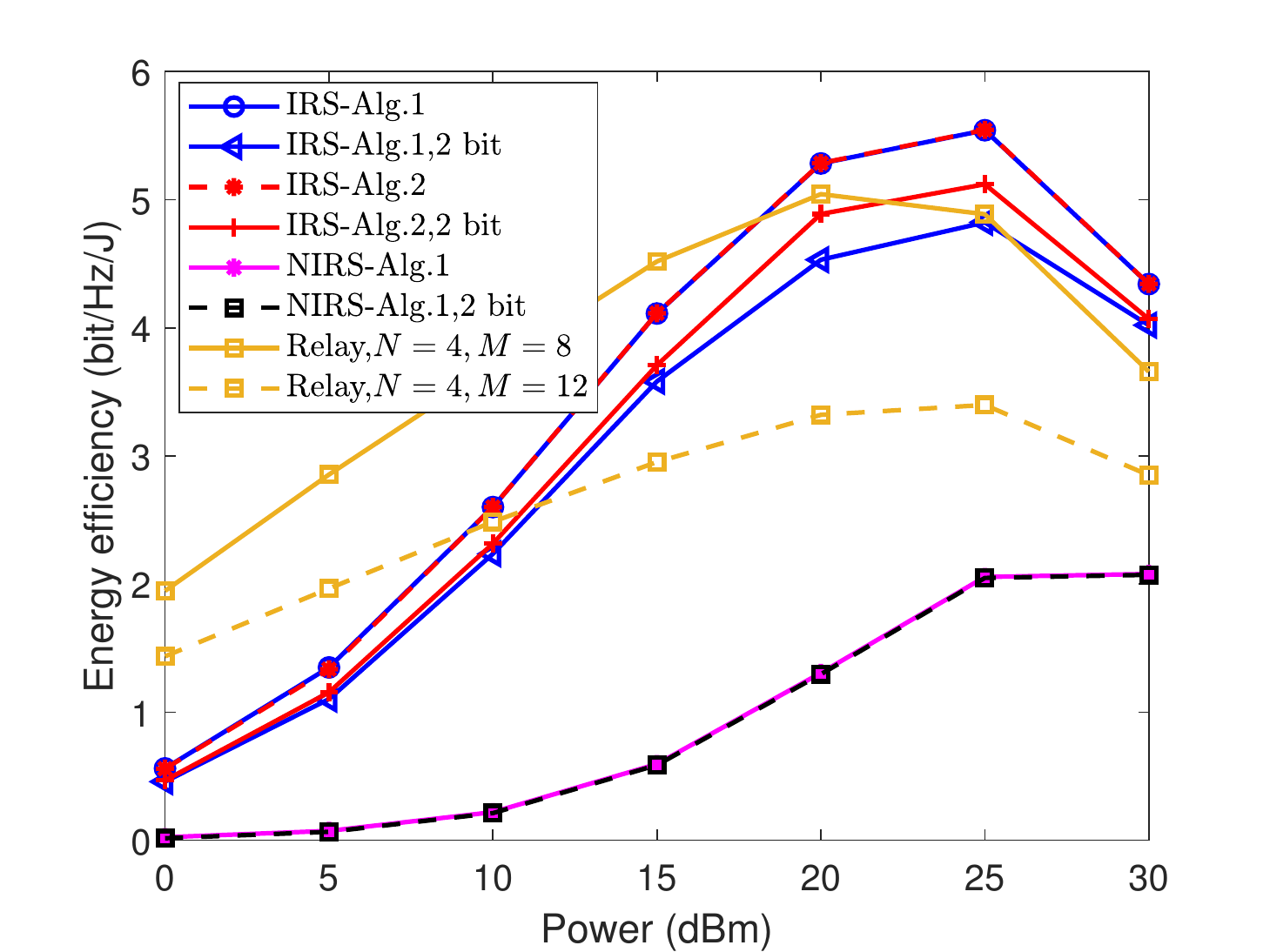}}

\centering \subfigure[CPU time versus transmit power ]{\includegraphics[width=3.4in,height=2.6in]{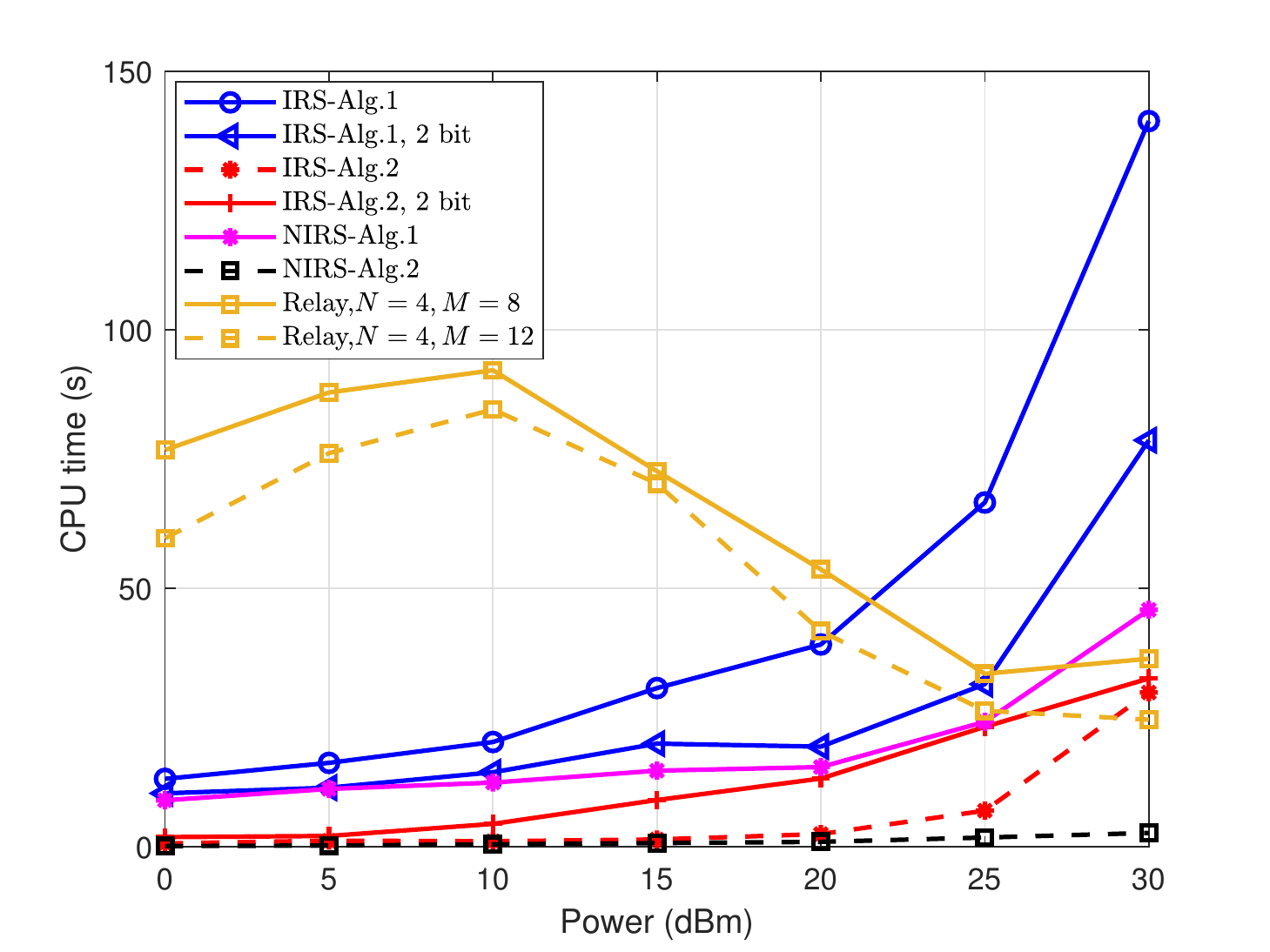}}

\caption{The sum rate, energy efficiency, and CPU time versus the transmit
power, when $N=4$, $N=16$ and $G=|\mathcal{K}_{g}|=2$.}
\label{rate_CPU_SNR} 
\end{figure}

Fig. \ref{rate_CPU_SNR} shows the sum rate, the energy efficiency,
and the corresponding CPU running time under different maximum transmit
power. The energy efficiency (bit/Hz/J) is defined as the ratio of
the sum rate to the power consumption, i.e., 
\[
EE=\frac{Sum\thinspace\thinspace Rate}{Power}\text{.}
\]
In the relay-aided system, we set $P_{\mathrm{T}}=P_{\mathrm{relay}}$.
The linear power consumption model is $Power=\eta(p_{\mathrm{T}}+p_{\mathrm{relay}})+NP_{t}+2MP_{r}$,
where $p_{\mathrm{T}}$ and $p_{\mathrm{relay}}$ are the practical
transmit power of the BS and the relay, respectively. Following \cite{shuguang2005},
we set the reciprocal of the power amplifier efficiency as $\eta=1.2$
and the circuit power consumption of the active antennas at the BS
and the relay as $P_{t}=P_{r}=200$ mW. In the IRS-aided system, we
adopt $Power=\eta(p_{\mathrm{T}}+p_{\mathrm{relay}})+NP_{t}+MP_{IRS}$,
where the circuit power consumption of the passive reflection elements
is set as $P_{IRS}=5$ mW \cite{emil2019}.

It can be seen in Fig. \ref{rate_CPU_SNR}(a) that the IRS structure
can obviously enhance the sum rate performance of the system without
consuming additional transmit power, comparing with the system without
the IRS structure. The performance loss of the `2 bit' phase shifter
generated by the proposed two algorithms is much small compared with
the continuous phase shifter cases. However, the relay-aided system
outperforms the IRS-aided one, which is reasonable due the fact that
the relay can amplify and forward the received signals by using the
relay transmit power $P_{\mathrm{relay}}$. The EE of the IRS-aided
system shown in Fig. \ref{rate_CPU_SNR}(b) is higher than the relay-aided
one at high transmit power. The reason behind this is twofold. On
the one hand, as $P_{\mathrm{T}}$ increases, the contribution of
the relay transmit power $P_{\mathrm{relay}}$ to the system sum rate
gain becomes less. On the other hand, the circuit power consumption
of the relay is relatively high. Another observation from Fig. \ref{rate_CPU_SNR}(b)
is that the EE of the relay system decreases with the number of the
active antennas deployed at the relay. From Fig. \ref{rate_CPU_SNR}(c),
we observe that Algorithm \ref{Algorithm-SOCP} is time-consuming
and the time required is unacceptable when $P_{\mathrm{T}}$ increases.
In addition, the computational complexity of the joint optimization
of the precoder and the relay beamforming is much higher than the
IRS case when $P_{\mathrm{T}}$ is less than 20 dBm due to the fact
that relay power constraint is complex. Finally, all the results obtained
from Fig. \ref{rate_CPU_SNR} verify the performance gains of the
IRS-aided system in terms of the EE and complexity.

\subsection{IRS Performance Analysis}

\begin{figure}
\centering \subfigure[Sum rate]{\includegraphics[width=3.4in,height=2.6in]{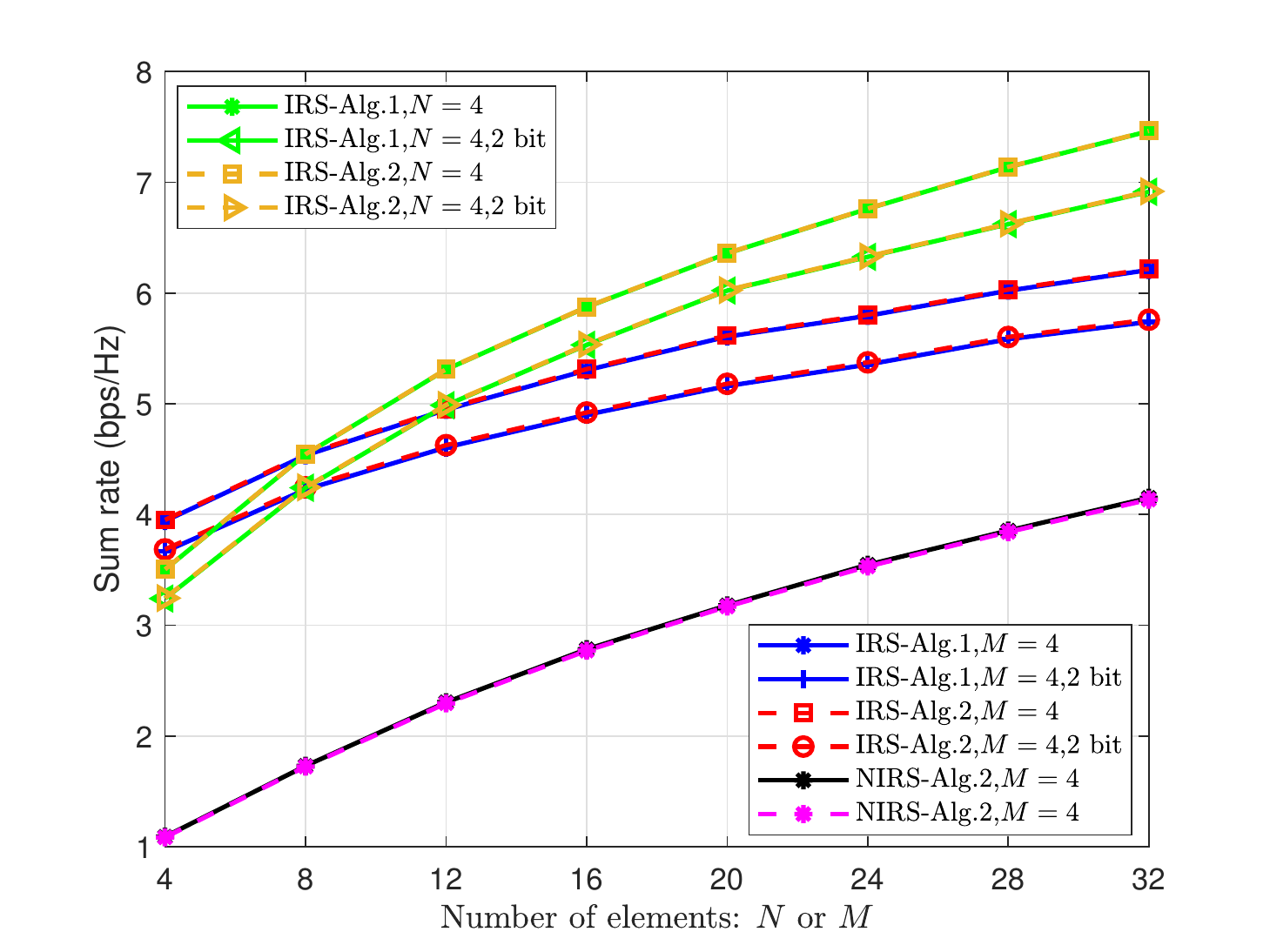}}

\centering \subfigure[Energy efficiency ]{\includegraphics[width=3.4in,height=2.6in]{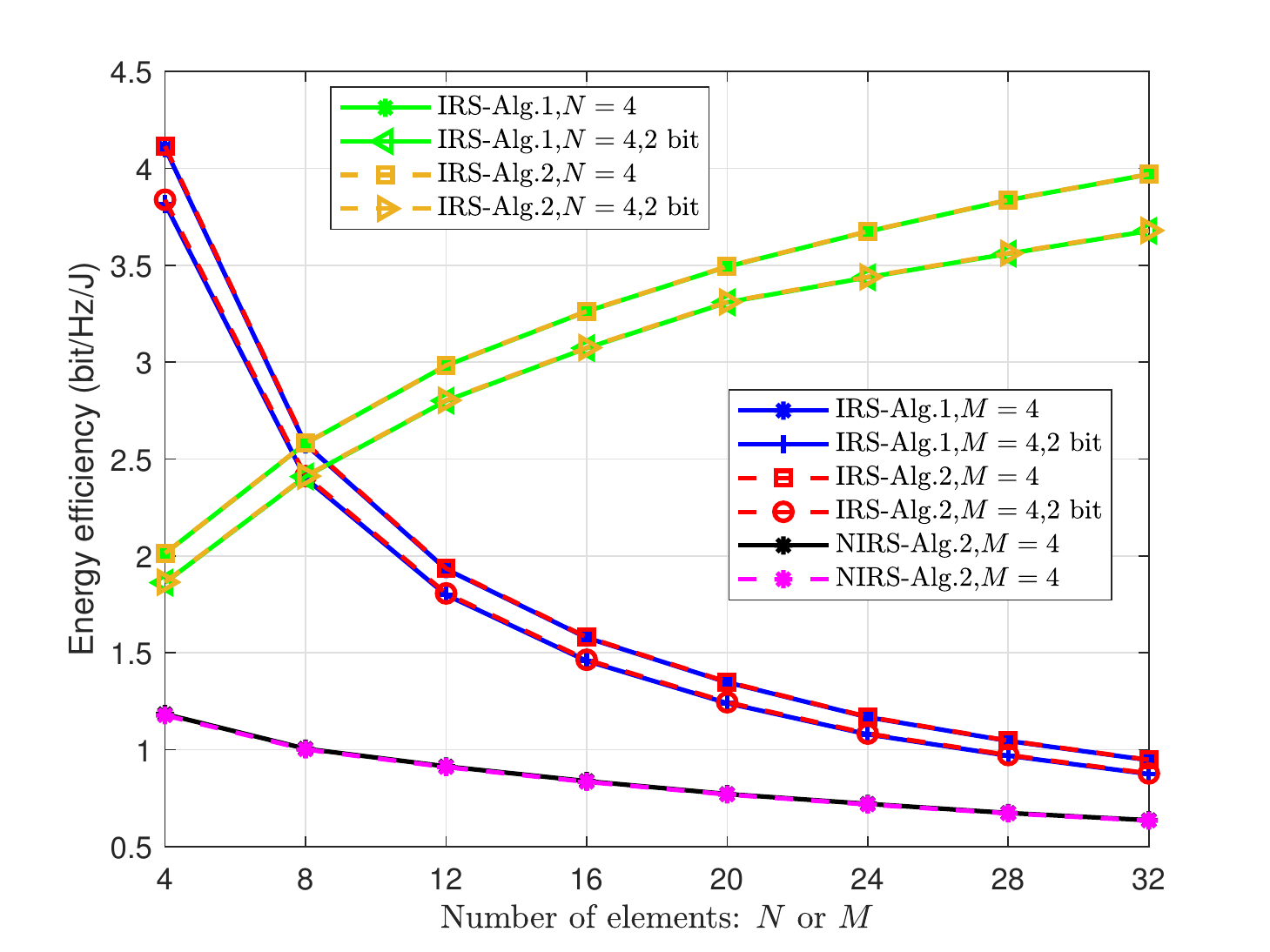}}

\caption{The sum rate versus the numbers of reflection elements at the IRS
$M$ or transmit antennas at the BS $N$, when $G=|\mathcal{K}_{g}|=2$
and $P_{\mathrm{T}}=20$ dBm.}
\label{rate_M} 
\end{figure}

It is of practical significance to compare the communication performance
of conventional large-scale antenna arrays deployed at the BS and
large-scale passive elements deployed at the IRS, since IRS is regarded
as an extension of massive MIMO antenna array. Fig. \ref{rate_M}
illustrates the sum rate and the EE performance versus the numbers
of antenna elements at the BS and reflection elements at the IRS when
$P_{\mathrm{T}}=20$ dBm. It is observed from Fig. \ref{rate_M}(a)
that significant gains can be achieved by the IRS scheme over that
without an IRS even when $M$ is as small as 4, and also that the
spectral efficiency performance gains achieved by increasing the number
of reflection elements are much higher than those achieved by increasing
the number of transmit antennas. In addition, in Fig. \ref{rate_M}(b),
it is more energy-efficient to deploy an IRS with passive elements
than installing active large-scale antenna array with energy-consuming
radio frequency chains and power amplifiers. The trend of EE decreasing
with the number of transmit antennas comes from the fact that the
circut energy consumption of more antennas outweighs the system sum
rate gain introduced by deploying more antennas. These simulation
results demonstrate that IRS technology is superior to traditional
massive MIMO in terms of spectral efficiency and energy efficiency.

\begin{figure}
\centering \includegraphics[width=3.4in,height=2.6in]{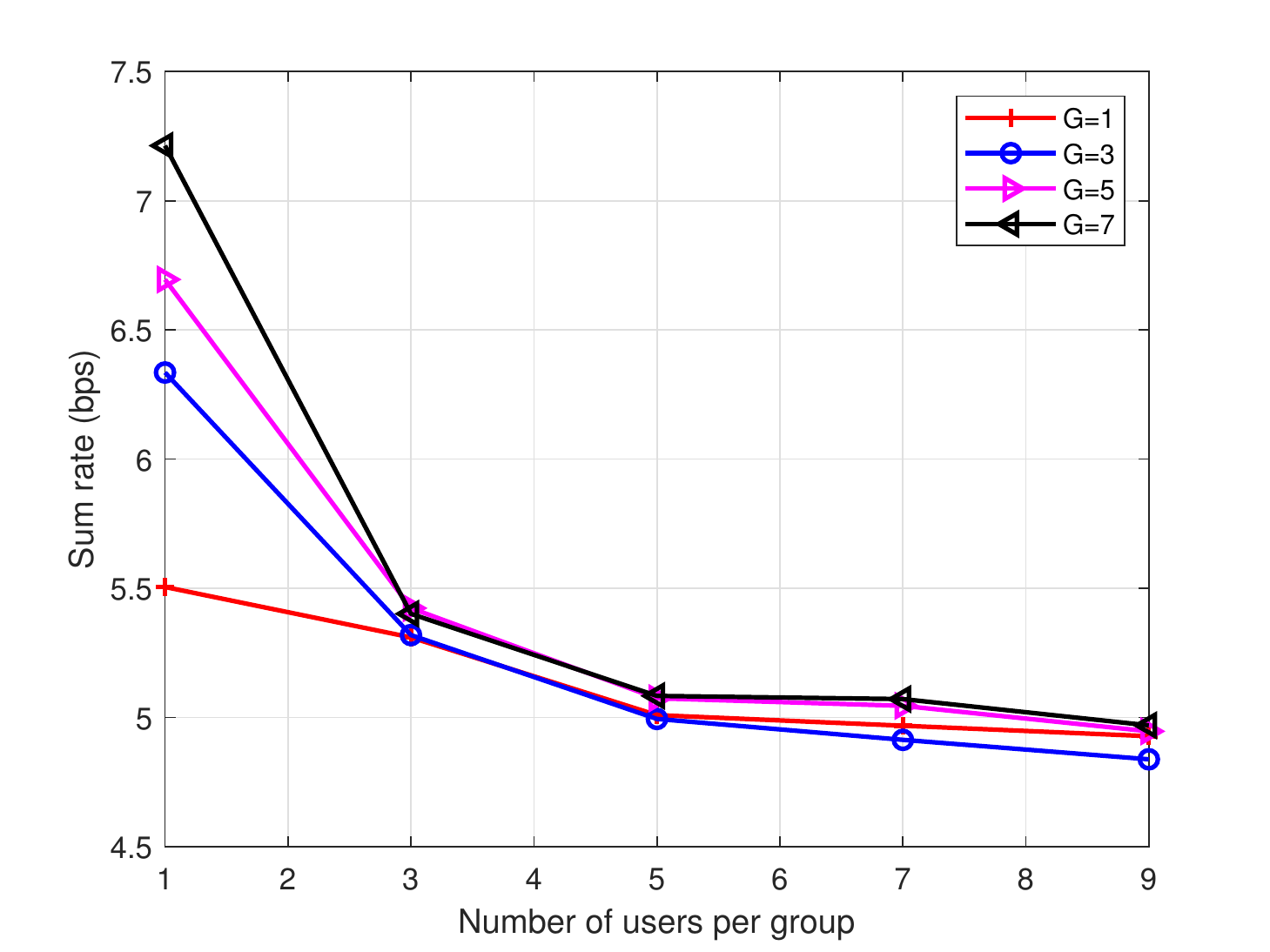}
\caption{The sum rate versus the number of users per group, when $N=4$, $N=16$,
and $P_{\mathrm{T}}=20$ dBm.}
\label{rate_users} 
\end{figure}

The above simulation results show that Algorithm \ref{Algorithm-MM}
requires less CPU time than Algorithm \ref{Algorithm-SOCP}. Hence,
we adopt Algorithm \ref{Algorithm-MM} to investigate the effect of
an IRS on the performance of a multicast communication system. Fig.
\ref{rate_users} illustrates the sum rate versus the number of users
per group for various numbers of groups. It can be observed from this
figure that the sum rate for all values of $G$ decreases with the
increase of the number of users per group. The reason is that the
data rate for each group is limited by the user with the worst channel
condition. With the increase of the number of users per group, the
channel gain for the worst user becomes smaller. 

Fig. \ref{rate_users} compares the effects of two improvements on
the performance limit, namely, increasing the number of antennas at
the BS and the number of reflection elements at the IRS, respectively.
When $|\mathcal{K}_{g}|=1$, the multicasting system reduces to a
unitcasting system, in which the transmit antennas outperform the
reflection elements in the aspect of suppressing multi-user interference.
While when $|\mathcal{K}_{g}|=3$, the sum rate of the system increases
slowly and tends to be stable with the increase of the number of multicasting
groups for a given number of antenna/reflection elements.

\begin{figure}
\centering \includegraphics[width=3.4in,height=2.6in]{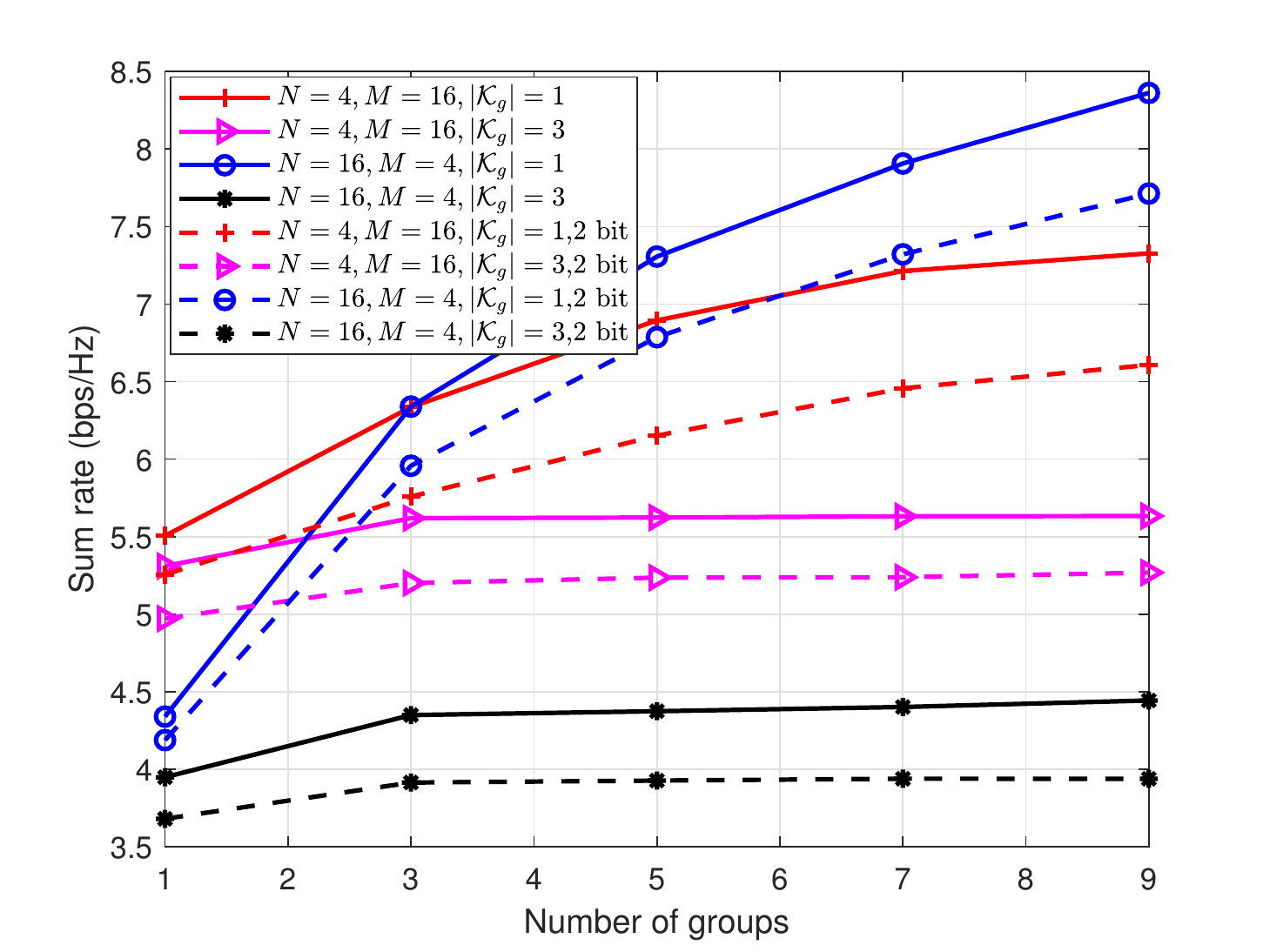}
\caption{The sum rate versus the number of groups, when $P_{\mathrm{T}}=20$
dBm.}
\label{rate_group} 
\end{figure}

\section{Conclusions}

In this work, we have shown the performance benefits of introducing
an IRS to the multigroup multicast systems. By carefully adjusting
the reflection coefficients at the IRS, the signal reflected by the
IRS can enhance the strength of the signal received by the user. We
investigate the sum rate maximization problem by joint optimization
of the precoding matrix at the BS and reflection coefficient vector
at the IRS, while guaranteeing the transmit power constraint and the
associated non-convex unit-modulus constraint at the IRS. Under the
MM algorithm framework, we derive the concave lower bound of the original
non-concave objective function, and then adopt alternating optimization
method to update variables in an alternating manner. Furthermore,
we proposed a low-complexity algorithm under the MM algorithm framework
in which there exists closed-form solutions at each iteration. Our
simulation results have demonstrated the significant spectral and
energy efficiency enhancement of the IRS in multigroup multicast systems
and that the proposed algorithm converges rapidly in terms of CPU
time.

\appendices{}

\section{The proof of Theorem \ref{lemma-1}\label{subsec:The-proof-of-1}}

We perform some equivalent transformations of the rate expression
(\ref{eq:Rate-k-1}) to show its hidden convexity, as follows 
\begin{align}
R_{k}\left(\mathbf{F},\mathbf{e}\right) & =\log_{2}\left(1+\frac{|\mathbf{e}^{\mathrm{H}}\mathbf{H}_{k}{\bf f}_{g}|^{2}}{\sum_{i\neq g}^{G}|\mathbf{e}^{\mathrm{H}}\mathbf{H}_{k}{\bf f}_{i}|^{2}+\sigma_{k}^{2}}\right)\nonumber \\
 & =\log_{2}\left(1+r_{k,-g}^{-1}|\mathbf{e}^{\mathrm{H}}\mathbf{H}_{k}{\bf f}_{g}|^{2}\right)\nonumber \\
 & =-\log_{2}\left(1-\left(r_{k,-g}+|\mathbf{e}^{\mathrm{H}}\mathbf{H}_{k}{\bf f}_{g}|^{2}\right)^{-1}|\mathbf{e}^{\mathrm{H}}\mathbf{H}_{k}{\bf f}_{g}|^{2}\right)\nonumber \\
 & =-\log_{2}\left(1-r_{k}^{-1}|t_{k}|^{2}\right),\label{eq:joint-convex}
\end{align}
where $t_{k}=\mathbf{e}^{\mathrm{H}}\mathbf{H}_{k}{\bf f}_{g}$, $r_{k}=r_{k,-g}+|t_{k}|^{2}$,
and $r_{k,-g}=\sum_{i\neq g}^{G}|\mathbf{e}^{\mathrm{H}}\mathbf{H}_{k}{\bf f}_{i}|^{2}+\sigma_{k}^{2}$.

Denoting $R_{k}(t_{k},r_{k})$ as the last equation expression of
$R_{k}\left(\mathbf{F},\mathbf{e}\right)$ in (\ref{eq:joint-convex}),
$R_{k}(t_{k},r_{k})$ is jointly convex in $\{t_{k},r_{k}\}$ \cite{wang2016},
thus its lower bound surrogate function could be obtained by the first-order
approximation, e.g., 
\begin{align}
 & R_{k}\left(t_{k},r_{k}\right)\nonumber \\
 & \geq R_{k}\left(t_{k}^{n},r_{k}^{n}\right)+\frac{\partial R_{k}}{\partial t_{k}}|_{t_{k}=t_{k}^{n}}\left(t_{k}-t_{k}^{n}\right)\nonumber \\
 & \ \ \ +\frac{\partial R_{k}}{\partial t_{k}^{*}}|_{t_{k}^{*}=t_{g}^{n,*}}\left(t_{k}^{*}-t_{k}^{n,*}\right)+\frac{\partial R_{k}}{\partial r_{k}}|_{r_{k}=r_{k}^{n}}\left(r_{k}-r_{k}^{n}\right)\nonumber \\
 & =R_{k}\left(t_{k}^{n},r_{k}^{n}\right)+2\mathit{\mathrm{Re}}\left\{ \frac{t_{k}^{n,*}(t_{k}-t_{k}^{n})}{r_{k}^{n}-|t_{k}^{n}|^{2}}\right\} -\frac{|t_{k}^{n}|^{2}(r_{k}-r_{k}^{n})}{r_{k}^{n}(r_{k}^{n}-|t_{k}^{n}|^{2})}\nonumber \\
 & =R_{k}\left(t_{k}^{n},r_{k}^{n}\right)+2\mathit{\mathrm{Re}}\left\{ \frac{t_{k}^{n,*}}{r_{k}^{n}-|t_{k}^{n}|^{2}}t_{k}\right\} \nonumber \\
 & \thinspace\thinspace\thinspace\thinspace\thinspace\thinspace\thinspace\thinspace\thinspace\thinspace-\frac{|t_{k}^{n}|^{2}}{r_{k}^{n}(r_{k}^{n}-|t_{k}^{n}|^{2})}r_{k}-\frac{|t_{k}^{n}|^{2}}{r_{k}^{n}-|t_{k}^{n}|^{2}}.\label{eq:surrogate-rate-1}
\end{align}

Undo $t_{k}=\mathbf{e}^{\mathrm{H}}\mathbf{H}_{k}{\bf f}_{g}$, $t_{k}^{n}=(\mathbf{e}^{n})^{\mathrm{H}}\mathbf{H}_{k}{\bf f}_{g}^{n}$,
$r_{k}=\sum_{i=1}^{G}|\mathbf{e}^{\mathrm{H}}\mathbf{H}_{k}{\bf f}_{i}|^{2}+\sigma_{k}^{2}$,
and $r_{k}^{n}=\sum_{i=1}^{G}|(\mathbf{e}^{n})^{\mathrm{H}}\mathbf{H}_{k}{\bf f}_{i}^{n}|^{2}+\sigma_{k}^{2}$,
and substitute them into the right hand side of the last equation
in (\ref{eq:surrogate-rate-1}), we have 
\begin{align}
R_{k}\left(\mathbf{F},\mathbf{e}\right) & \geq R_{k}\left(\mathbf{F}^{n},\mathbf{e}^{n}\right)+2\textrm{\ensuremath{\mathrm{Re}}}\left\{ a_{k}\mathbf{e}^{\mathrm{H}}\mathbf{H}_{k}{\bf f}_{g}\right\} -\frac{|t_{k}^{n}|^{2}}{r_{k}^{n}-|t_{k}^{n}|^{2}}\nonumber \\
 & \thinspace\thinspace\thinspace\thinspace\thinspace\thinspace\thinspace-b_{k}\sum_{i=1}^{G}|\mathbf{e}^{\mathrm{H}}\mathbf{H}_{k}{\bf f}_{i}|^{2}-b_{k}\sigma_{k}^{2}\nonumber \\
 & =\textrm{const}_{k}+2\textrm{\ensuremath{\mathrm{Re}}}\left\{ a_{k}\mathbf{e}^{\mathrm{H}}\mathbf{H}_{k}{\bf f}_{g}\right\} -b_{k}\sum_{i=1}^{G}|\mathbf{e}^{\mathrm{H}}\mathbf{H}_{k}{\bf f}_{i}|^{2}\nonumber \\
 & =\widetilde{R}_{k}\left(\mathbf{F},\mathbf{e}\right).\label{eq:surrogate-rate-2}
\end{align}

Hence, the proof is complete.

\section{The proof of Theorem \ref{Theorem-socp-kkt}\label{subsec:The-proof-of-5}}

The monotonic property of the objective function value sequence $\{F(\mathbf{F}^{n},\mathbf{e}^{n})\}$
of Algorithm \ref{Algorithm-SOCP} can be guaranteed by (\ref{eq:dfd}).
In addition, the sequence $\{\mathbf{F}^{n},\mathbf{e}^{n}\}$ generated
at each iteration of Algorithm \ref{Algorithm-SOCP} converges to
a stable point as $n\rightarrow\infty$ because $\mathbf{F}^{n}$
and $\mathbf{e}^{n}$ are bounded in their feasible sets $\mathcal{S}_{F}$
and $\mathcal{S}_{e}$, respectively \cite{nasir2017secrecy}. Denote
by $\{\mathbf{F}^{o},\mathbf{e}^{o}\}$ the converged solution. In
the following, we prove that $\{\mathbf{F}^{o},\mathbf{e}^{o}\}$
is the KKT point based on the fact that all the locally optimal solutions
(including the globally optimal solution) of a nonconvex optimization
problem should satisfy the KKT optimality conditions \cite{book-convex}.

Firstly, the Lagrangian of Problem (\ref{eq:Problem-F-socp}) is given
by 
\begin{align*}
 & \mathcal{L}\text{(\ensuremath{\mathbf{F}},\ensuremath{\boldsymbol{\gamma}},\ensuremath{\boldsymbol{\lambda}^{(1)}},\ensuremath{\lambda^{(2)}})}\\
= & \sum_{g=1}^{G}\gamma_{g}-\sum_{g=1}^{G}\sum_{k\in\mathcal{K}_{g}}\lambda_{k}^{(1)}(\gamma_{g}-\widetilde{R}_{k}\left(\mathbf{F},\mathbf{e}^{o}|\mathbf{F}^{o},\mathbf{e}^{o}\right))\\
 & -\lambda^{(2)}(\mathrm{Tr}\left[\mathbf{F}^{\mathrm{H}}\mathbf{F}\right]-P_{\mathrm{T}})
\end{align*}
where $\boldsymbol{\lambda}^{(1)}=[\lambda_{1}^{(1)},...,\lambda_{K}^{(1)}]$
and $\lambda^{(2)}$ are the dual variables. Since $\mathbf{F}^{o}$
is the globally optimal solution of Problem (\ref{eq:Problem-F-socp}),
there must exist a $\boldsymbol{\lambda}^{(1),o}$ and $\lambda^{(2),o}$
satisfying the following partial KKT conditions: 
\begin{align}
 & \sum_{g=1}^{G}\sum_{k\in\mathcal{K}_{g}}\lambda_{k}^{(1),o}\nabla_{\mathbf{F}^{*}}\widetilde{R}_{k}\left(\mathbf{F},\mathbf{e}^{o}|\mathbf{F}^{o},\mathbf{e}^{o}\right)|_{\mathbf{F}=\mathbf{F}^{o}}-\lambda^{(2),o}\mathbf{F}^{o}=\mathbf{0},\label{eq:KKT-f-1}\\
 & \lambda_{k}^{(1),o}(\gamma_{g}-\widetilde{R}_{k}\left(\mathbf{F}^{o},\mathbf{e}^{o}|\mathbf{F}^{o},\mathbf{e}^{o}\right))=0,\forall k\in\mathcal{K}_{g},\forall g\in\mathcal{G},\label{eq:KKT-f-2}\\
 & \lambda^{(2),o}(\mathrm{Tr}\left[\mathbf{F}^{\mathrm{H},o}\mathbf{F}^{o}\right]-P_{\mathrm{T}})=0.\label{eq:KKT-f-3}
\end{align}
According to the conditions (A1) and (A3), we have 
\begin{align}
\widetilde{R}_{k}\left(\mathbf{F}^{o},\mathbf{e}^{o}|\mathbf{F}^{o},\mathbf{e}^{o}\right) & =R_{k}\left(\mathbf{F}^{o},\mathbf{e}^{o}\right),\label{eq:A1-f}\\
\nabla_{\mathbf{F}^{*}}\widetilde{R}_{k}\left(\mathbf{F},\mathbf{e}^{o}|\mathbf{F}^{o},\mathbf{e}^{o}\right)|_{\mathbf{F}=\mathbf{F}^{o}} & =\nabla_{\mathbf{F}^{*}}R_{k}\left(\mathbf{F},\mathbf{e}^{o}\right)|_{\mathbf{F}=\mathbf{F}^{o}}.\label{eq:A3-f}
\end{align}
By substituting (\ref{eq:A3-f}) and (\ref{eq:A1-f}) into (\ref{eq:KKT-f-1})
and (\ref{eq:KKT-f-2}) respectively, we arrive at 
\begin{align}
 & \sum_{g=1}^{G}\sum_{k\in\mathcal{K}_{g}}\lambda_{k}^{(1),o}\nabla_{\mathbf{F}^{*}}R_{k}\left(\mathbf{F},\mathbf{e}^{o}\right)|_{\mathbf{F}=\mathbf{F}^{o}}-\lambda^{(2),o}\mathbf{F}^{o}=\mathbf{0},\label{eq:KKT-f-4}\\
 & \lambda_{k}^{(1),o}(\gamma_{g}-R_{k}\left(\mathbf{F}^{o},\mathbf{e}^{o}\right))=0,\forall k\in\mathcal{K}_{g},\forall g\in\mathcal{G}.\label{eq:KKT-f-5}
\end{align}

Then, $\mathbf{e}^{o}$ is the locally optimal solution of Problem
(\ref{eq:Problem-e-2}) and satisfies the following KKT conditions:
\begin{align}
 & \sum_{g=1}^{G}\sum_{k\in\mathcal{K}_{g}}\xi_{k}^{(1),o}\nabla_{\mathbf{e}^{*}}\widetilde{R}_{k}\left(\mathbf{F}^{o},\mathbf{e}|\mathbf{F}^{o},\mathbf{e}^{o}\right)|_{\mathbf{e}=\mathbf{e}^{o}}-\nonumber \\
 & \sum_{m=1}^{M}\xi_{m}^{(2),o}(\nabla_{\mathbf{e}^{*}}|e_{m}|)|_{\mathbf{e}=\mathbf{e}^{o}}-\xi_{M+1}^{(2),o}(\nabla_{\mathbf{e}^{*}}e_{M+1})|_{\mathbf{e}=\mathbf{e}^{o}}=\mathbf{0},\label{eq:KKT-e-1}\\
 & \xi_{k}^{(1),o}(\kappa_{g}-\widetilde{R}_{k}\left(\mathbf{F}^{o},\mathbf{e}^{o}|\mathbf{F}^{o},\mathbf{e}^{o}\right))=0,\forall k\in\mathcal{K}_{g},\forall g\in\mathcal{G},\label{eq:KKT-e-2}\\
 & \xi_{m}^{(2),o}(|e_{m}^{o}|-1)=0,1\leq m\leq M,\xi_{M+1}^{(2),o}(e_{M+1}^{o}-1)=0,\label{eq:KKT-e-3}
\end{align}
where $\boldsymbol{\xi}^{(1),o}=[\xi_{1}^{(1),o},...,\xi_{K}^{(1),o}]$
and $\xi^{(2),o}$ are the optimal Lagrange multipliers.

Furthermore, it can be readily checked that 
\begin{align}
\nabla_{\mathbf{e}^{*}}\widetilde{R}_{k}\left(\mathbf{F}^{o},\mathbf{e}|\mathbf{F}^{o},\mathbf{e}^{o}\right)|_{\mathbf{e}=\mathbf{e}^{o}} & =\nabla_{\mathbf{e}^{*}}R_{k}\left(\mathbf{F}^{o},\mathbf{e}\right)|_{\mathbf{e}=\mathbf{e}^{o}}.\label{eq:A3-e}
\end{align}
By substituting (\ref{eq:A3-e}) into (\ref{eq:KKT-e-1}), we arrive
at 
\begin{align}
 & \sum_{g=1}^{G}\sum_{k\in\mathcal{K}_{g}}\xi_{k}^{(1),o}\nabla_{\mathbf{e}^{*}}R_{k}\left(\mathbf{F}^{o},\mathbf{e}\right)|_{\mathbf{e}=\mathbf{e}^{o}}-\xi_{M+1}^{(2),o}(\nabla_{\mathbf{e}^{*}}e_{M+1})|_{\mathbf{e}=\mathbf{e}^{o}}\nonumber \\
 & -\sum_{m=1}^{M}\xi_{m}^{(2),o}(\nabla_{\mathbf{e}^{*}}|e_{m}|)|_{\mathbf{e}=\mathbf{e}^{o}}=\mathbf{0},\label{eq:KKT-e-4}
\end{align}

Now, we move to Problem (\ref{eq:Problem-original}). The general
equivalent problem of the max-min Problem (\ref{eq:Problem-original})
is given by 
\begin{align}
\mathop{\max}\limits _{\mathbf{F},\mathbf{e},\mathbf{r}} & \;\;\sum_{g=1}^{G}r_{g}\nonumber \\
{\rm s.t.} & \;\;\mathbf{F}\in\mathcal{S}_{F},\mathbf{e}\in\mathcal{S}_{e}\nonumber \\
 & \;\;R_{k}\left(\mathbf{F},\mathbf{e}\right)\geq r_{g},\forall k\in\mathcal{K}_{g},\forall g\in\mathcal{G}.\label{eq:original-2}
\end{align}
where $\mathbf{r}=[r_{1},...,r_{G}]^{\mathrm{T}}$ are auxiliary variables.
It can be readily verified that the set of equations (\ref{eq:KKT-f-4}),
(\ref{eq:KKT-e-4}), (\ref{eq:KKT-f-5}), (\ref{eq:KKT-f-3}), and
(\ref{eq:KKT-e-3}) constitute exactly the KKT conditions of Problem
(\ref{eq:original-2}).

Hence, the proof is complete.

\section{The proof of Theorem \ref{Theorem-1}\label{subsec:The-proof-of-2}}

Since $f_{g}\left(\mathbf{F}\right)$ is twice differentiable and
concave, we propose a quadratic surrogate function to minorize $f_{g}\left(\mathbf{F}\right)$,
as follows 
\begin{align}
f_{g}(\mathbf{F}) & \geq f_{g}(\mathbf{F}^{n})+2\textrm{\ensuremath{\mathrm{Re}}}\left\{ \mathrm{\mathrm{Tr}}\left[\mathbf{D}_{g}^{\mathrm{H}}(\mathbf{F}-\mathbf{F}^{n})\right]\right\} \nonumber \\
 & \thinspace\thinspace\thinspace\thinspace\thinspace\thinspace+\mathrm{Tr}\left[(\mathbf{F}-\mathbf{F}^{n})^{\mathrm{H}}\mathbf{M}_{g}(\mathbf{F}-\mathbf{F}^{n})\right]\label{eq:quadratic-f}
\end{align}
where matrices $\mathbf{D}_{g}\in\mathbb{C}^{N\times N}$ and $\mathbf{M}_{g}\in\mathbb{C}^{N\times N}$
are determined to satisfy conditions (A1)-(A4).

Note that (A1) and (A4) are already satisfied. Then we prove that
condition (A3) also holds. Let $\widetilde{\mathbf{F}}$ be a matrix
belonging to $\mathcal{S}_{F}$. The directional derivative of the
right hand side of (\ref{eq:quadratic-f}) at $\mathbf{F}^{n}$ with
direction $\widetilde{\mathbf{F}}-\mathbf{F}^{n}$ is given by: 
\begin{equation}
2\textrm{Re}\left\{ \mathrm{\mathrm{Tr}}\left[\mathbf{D}_{g}^{\mathrm{H}}(\widetilde{\mathbf{F}}-\mathbf{F}^{n})\right]\right\} .\label{eq:First-left-f}
\end{equation}

The directional derivative of $f_{g}(\mathbf{F})$ is 
\begin{equation}
2\textrm{\textrm{\ensuremath{\mathrm{Re}}}}\left\{ \mathrm{\mathrm{Tr}}\left[\sum_{k\in\mathcal{K}_{g}}g_{k}(\mathbf{F}^{n})(\mathbf{C}_{k}^{\mathrm{H}}-(\mathbf{F}^{n})^{\mathrm{H}}\mathbf{B}_{k})(\widetilde{\mathbf{F}}-\mathbf{F}^{n})\right]\right\} ,\label{eq:First-right-f}
\end{equation}
where $g_{k}(\mathbf{F}^{n})$ is defined in (\ref{g_k_f}).

In order to satisfy condition (A3), the two directional derivatives
(\ref{eq:First-left-f}) and (\ref{eq:First-right-f}) must be equal,
which means 
\begin{equation}
\mathbf{D}_{g}=\sum_{k\in\mathcal{K}_{g}}g_{k}(\mathbf{F}^{n})(\mathbf{C}_{k}-\mathbf{B}_{k}^{\mathrm{H}}\mathbf{F}^{n}).
\end{equation}

Now we proceed to prove that condition (A2) also holds. If surrogate
function $\widetilde{f}_{g}(\mathbf{F}|\mathbf{F}^{n})$ is a lower
bound for each linear cut in any direction, condition (A2) could be
satisfied. Let $\mathbf{F}=\mathbf{F}^{n}+\gamma(\widetilde{\mathbf{F}}-\mathbf{F}^{n}),\forall\gamma\in[0,1]$.
Then, it suffices to show 
\begin{align}
f_{g}(\mathbf{F}^{n}+\gamma(\widetilde{\mathbf{F}}-\mathbf{F}^{n})) & \geq f_{g}(\mathbf{F}^{n})+2\gamma\textrm{\ensuremath{\mathrm{Re}}}\left\{ \mathrm{\mathrm{Tr}}\left[\mathbf{D}_{g}^{\mathrm{H}}(\widetilde{\mathbf{F}}-\mathbf{F}^{n})\right]\right\} \nonumber \\
+\gamma^{2} & \mathrm{\mathrm{Tr}}\left[(\widetilde{\mathbf{F}}-\mathbf{F}^{n})^{\mathrm{H}}\mathbf{M}_{g}(\widetilde{\mathbf{F}}-\mathbf{F}^{n})\right],\label{eq:quadratic-f-A2}
\end{align}

Let us define $L_{g}(\gamma)=f_{g}(\mathbf{F}^{n}+\gamma(\widetilde{\mathbf{F}}-\mathbf{F}^{n})),$and
$l_{k}(\gamma)=\widetilde{R}_{k}(\mathbf{F}^{n}+\gamma(\widetilde{\mathbf{F}}-\mathbf{F}^{n})).$
Now, a sufficient condition for (\ref{eq:quadratic-f-A2}) to hold
is that the second derivative of the right hand side of (\ref{eq:quadratic-f-A2})
is lower than or equal to the second derivative of the left hand side
of (\ref{eq:quadratic-f-A2}) for $\forall\gamma\in[0,1]$ and $\forall\widetilde{\mathbf{F}},\forall\mathbf{F}^{n}\in\mathcal{S}_{F}$,
which is formulated as follows 
\begin{equation}
\frac{\partial^{2}L_{g}(\gamma)}{\partial\gamma^{2}}\geq2\mathrm{\mathrm{Tr}}\left[(\widetilde{\mathbf{F}}-\mathbf{F}^{n})^{\mathrm{H}}\mathbf{M}_{g}(\widetilde{\mathbf{F}}-\mathbf{F}^{n})\right].\label{eq:second-inequation}
\end{equation}

In order to calculate the left hand side of (\ref{eq:second-inequation}),
we first calculate the first-order derivative, as follows 
\begin{align}
\frac{\partial L_{g}(\gamma)}{\partial\gamma} & =\sum_{k\in\mathcal{K}_{g}}g_{k}(\gamma)\nabla_{\gamma}l_{k}(\gamma),
\end{align}
where 
\begin{align*}
g_{k}(\gamma) & =\frac{\mathrm{exp}\left\{ -\mu_{g}l_{k}(\gamma)\right\} }{\sum_{k\in\mathcal{K}_{g}}\mathrm{exp}\left\{ -\mu_{g}l_{k}(\gamma)\right\} },k\in\mathcal{K}_{g},\\
\nabla_{\gamma}l_{k}(\gamma) & =2\textrm{\ensuremath{\mathrm{Re}}}\Bigl\{\mathrm{\mathrm{Tr}}\left[\mathbf{C}_{k}^{\mathrm{H}}(\widetilde{\mathbf{F}}-\mathbf{F}^{n})\right]\\
 & \ \ \ -\mathrm{Tr}\left[(\mathbf{F}^{n}+\gamma(\widetilde{\mathbf{F}}-\mathbf{F}^{n}))^{\mathrm{H}}\mathbf{B}_{k}(\widetilde{\mathbf{F}}-\mathbf{F}^{n})\right]\Bigr\}\\
 & =2\textrm{\ensuremath{\mathrm{Re}}}\left\{ \mathrm{\mathrm{Tr}}\left[\mathbf{Q}_{k}^{\mathrm{H}}(\widetilde{\mathbf{F}}-\mathbf{F}^{n})\right]\right\} \\
 & =2\textrm{\ensuremath{\mathrm{Re}}}\left\{ \mathbf{q}_{k}^{\mathrm{H}}\mathbf{f}\right\} ,\\
\mathbf{Q}_{k}^{\mathrm{H}} & =\mathbf{C}_{k}^{\mathrm{H}}-(\mathbf{F}^{n}+\gamma(\widetilde{\mathbf{F}}-\mathbf{F}^{n}))^{\mathrm{H}}\mathbf{B}_{k},\\
\mathbf{q}_{k} & =\mathrm{vec}(\mathbf{Q}_{k}),\\
\mathbf{f} & =\mathrm{vec}(\widetilde{\mathbf{F}}-\mathbf{F}^{n}),
\end{align*}
Then, the second-order derivative is derived as 
\begin{align}
 & \frac{\partial^{2}L_{g}(\gamma)}{\partial\gamma^{2}}\nonumber \\
 & =\sum_{k\in\mathcal{K}_{g}}\left(g_{k}(\gamma)\nabla_{\gamma}^{2}l_{k}(\gamma)-\mu_{g}g_{k}(\gamma)\nabla_{\gamma}l_{k}(\gamma)\left(\nabla_{\gamma}l_{k}(\gamma)\right)^{\mathrm{T}}\right)\nonumber \\
 & +\mu_{g}\left(\sum_{k\in\mathcal{K}_{g}}g_{k}(\gamma)\nabla_{\gamma}l_{k}(\gamma)\right)\left(\sum_{k\in\mathcal{K}_{g}}g_{k}(\gamma)\nabla_{\gamma}l_{k}(\gamma)\right)^{\mathrm{T}},\label{eq:second-derivative-right}
\end{align}
where 
\begin{align*}
\nabla_{\gamma}^{2}l_{k}(\gamma) & =-2\mathrm{\mathrm{Tr}}\left[(\widetilde{\mathbf{F}}-\mathbf{F}^{n})^{\mathrm{H}}\mathbf{B}_{k}(\widetilde{\mathbf{F}}-\mathbf{F}^{n})\right]\\
 & =-2\mathbf{f}{}^{\mathrm{H}}(\mathbf{I}\otimes\mathbf{B}_{k})\mathbf{f}.
\end{align*}

We reformulate $\frac{\partial^{2}L_{g}(\gamma)}{\partial\gamma^{2}}$
in (\ref{eq:second-derivative-right}) into a quadratic form of $\mathbf{f}$,
as follows 
\begin{align*}
\frac{\partial^{2}L_{g}(\gamma)}{\partial\gamma^{2}} & =\left[\begin{array}{cc}
\mathbf{f}{}^{\mathrm{H}} & \mathbf{f}{}^{\mathrm{T}}\end{array}\right]\boldsymbol{\Phi}\left[\begin{array}{c}
\mathbf{f}\\
\mathbf{f}{}^{*}
\end{array}\right],
\end{align*}
where $\boldsymbol{\Phi}$ is given in (\ref{Phi}).

\begin{figure*}
\begin{align}
\boldsymbol{\Phi}_{g} & =\sum_{k\in\mathcal{K}_{g}}\left(g_{k}(\gamma)\left[\begin{array}{cc}
-\mathbf{I}\otimes\mathbf{B}_{k} & \mathbf{0}\\
\mathbf{0} & -\mathbf{I}\otimes\mathbf{B}_{k}^{\mathrm{T}}
\end{array}\right]-\mu_{g}g_{k}(\gamma)\left[\begin{array}{c}
\mathbf{q}_{k}\\
\mathbf{q}_{k}^{\mathrm{*}}
\end{array}\right]\left[\begin{array}{c}
\mathbf{q}_{k}\\
\mathbf{q}_{k}^{\mathrm{*}}
\end{array}\right]^{\mathrm{H}}\right)+\mu_{g}\left[\begin{array}{c}
\sum_{k\in\mathcal{K}_{g}}g_{k}(\gamma)\mathbf{q}_{k}\\
\sum_{k\in\mathcal{K}_{g}}g_{k}(\gamma)\mathbf{q}_{k}^{\mathrm{*}}
\end{array}\right]\left[\begin{array}{c}
\sum_{k\in\mathcal{K}_{g}}g_{k}(\gamma)\mathbf{q}_{k}\\
\sum_{k\in\mathcal{K}_{g}}g_{k}(\gamma)\mathbf{q}_{k}^{\mathrm{*}}
\end{array}\right]^{\mathrm{H}}\text{.}\label{Phi}
\end{align}
\hrule 
\end{figure*}

We also manipulate the right hand side of (\ref{eq:second-inequation})
into a quadratic form of $\mathbf{f}$ by using vectorization operation
$\mathrm{Tr}[\mathbf{A}^{\mathrm{T}}\mathbf{B}\mathbf{C}]=\mathrm{vec}^{\mathrm{T}}(\mathbf{A})(\mathbf{I}\otimes\mathbf{B})\mathrm{vec}(\mathbf{C})$
\cite{maher1999handbook}, as follows 
\begin{align*}
 & 2\mathrm{\mathrm{Tr}}\left[(\widetilde{\mathbf{F}}-\mathbf{F}^{n})^{\mathrm{H}}\mathbf{M}_{g}(\widetilde{\mathbf{F}}-\mathbf{F}^{n})\right]\\
 & =\left[\begin{array}{cc}
\mathbf{f}{}^{\mathrm{H}} & \mathbf{f}{}^{\mathrm{T}}\end{array}\right]\left[\begin{array}{cc}
\mathbf{I}\otimes\mathbf{M}_{g} & \mathbf{0}\\
\mathbf{0} & \mathbf{I}\otimes\mathbf{M}_{g}^{\mathrm{T}}
\end{array}\right]\left[\begin{array}{c}
\mathbf{f}\\
\mathbf{f}{}^{*}
\end{array}\right].
\end{align*}
Then, (\ref{eq:second-inequation}) is equivalent to 
\begin{align*}
 & \left[\begin{array}{cc}
\mathbf{f}{}^{\mathrm{H}} & \mathbf{f}{}^{\mathrm{T}}\end{array}\right]\boldsymbol{\Phi}_{g}\left[\begin{array}{c}
\mathbf{f}\\
\mathbf{f}{}^{*}
\end{array}\right]\\
 & \geq\left[\begin{array}{cc}
\mathbf{f}{}^{\mathrm{H}} & \mathbf{f}{}^{\mathrm{T}}\end{array}\right]\left[\begin{array}{cc}
\mathbf{I}\otimes\mathbf{M}_{g} & \mathbf{0}\\
\mathbf{0} & \mathbf{I}\otimes\mathbf{M}_{g}^{\mathrm{T}}
\end{array}\right]\left[\begin{array}{c}
\mathbf{f}\\
\mathbf{f}{}^{*}
\end{array}\right],
\end{align*}
where we need to find an $\mathbf{M}_{g}$ that satisfies 
\[
\boldsymbol{\Phi}_{g}\succeq\left[\begin{array}{cc}
\mathbf{I}\otimes\mathbf{M}_{g} & \mathbf{0}\\
\mathbf{0} & \mathbf{I}\otimes\mathbf{M}_{g}^{\mathrm{T}}
\end{array}\right].
\]
For convenience, we choose $\mathbf{M}_{g}=\alpha_{g}\mathbf{I}=\lambda_{\mathrm{min}}\left(\boldsymbol{\Phi}_{g}\right)\mathbf{I}$.
Finally, (\ref{eq:quadratic-f}) is equivalent to 
\begin{align}
f_{g}(\mathbf{F}) & \geq f_{g}(\mathbf{F}^{n})+2\textrm{\ensuremath{\mathrm{Re}}}\left\{ \mathrm{Tr}\left[\mathbf{D}_{g}^{\mathrm{H}}(\mathbf{F}-\mathbf{F}^{n})\right]\right\} \nonumber \\
 & \thinspace\thinspace\thinspace\thinspace\thinspace\thinspace+\alpha_{g}\mathrm{\mathrm{Tr}}\left[(\mathbf{F}-\mathbf{F}^{n})^{\mathrm{H}}(\mathbf{F}-\mathbf{F}^{n})\right]\nonumber \\
 & =2\textrm{\ensuremath{\mathrm{Re}}}\left\{ \mathrm{\mathrm{Tr}}\left[\mathbf{U}_{g}^{\mathrm{H}}\mathbf{F}\right]\right\} +\alpha_{g}\mathrm{Tr}\left[\mathbf{F}^{\mathrm{H}}\mathbf{F}\right]+\textrm{consF}_{g}\label{eq:surrogate-f-final}
\end{align}
where $\mathbf{U}_{g}$ and $\textrm{consF}_{g}$ are given in (\ref{eq:U})
and (\ref{constant-f}), respectively. $\alpha_{g}$ in (\ref{alpha})
is difficult to obtain for the complex expression of $\boldsymbol{\Phi}_{g}$.
In the following, we proceed to obtain the value of $\alpha_{g}$.

The following inequalities and equalities will be used later:

(B1): \cite{maher1999handbook} $\mathbf{A}$ and $\mathbf{B}$ are
Hermitian matrices: $\lambda_{\mathrm{min}}(\mathbf{A})+\lambda_{\mathrm{min}}(\mathbf{B})\leq\lambda_{\mathrm{min}}(\mathbf{A}+\mathbf{B}).$

(B2): \cite{maher1999handbook} $\mathbf{A}$ is rank one: $\lambda_{\mathrm{max}}(\mathbf{A})=\mathrm{\mathrm{Tr}}\left[\mathbf{A}\right],\lambda_{\mathrm{min}}(\mathbf{A})=0$.

(B3): (Theorem 30 in \cite{book-Matrix}) $a_{k}$ and $b_{k}$ are
positive: $\sum_{k=1}^{K}a_{k}b_{k}\leq\mathrm{max}_{k=1}^{K}\left\{ b_{k}\right\} $,
if $\sum_{k=1}^{K}a_{k}=1$.

(B4): \cite{maher1999handbook} $\mathbf{A}$ is positive semidifinite
with maximum eigenvalue $\lambda_{\mathrm{max}}(\mathbf{A})$ and
$\mathbf{B}$ is positive semidifinite: $\mathrm{\mathrm{Tr}}\left[\mathbf{A}\mathbf{B}\right]\leq\lambda_{\mathrm{max}}(\mathbf{A})\mathrm{\mathrm{Tr}}\left[\mathbf{B}\right]$.

$\boldsymbol{\Phi}_{g}$ is complex and cannot be determined by a
constant, thus we use (A1)-(A4) to find its lower bound shown in (\ref{eq:eig-1}).
\begin{figure*}
\begin{align}
\lambda_{\mathrm{min}}\left(\boldsymbol{\Phi}_{g}\right) & \overset{\mathrm{(B1)}}{\geq}-\sum_{k\in\mathcal{K}_{g}}g_{k}(\gamma)\lambda_{\mathrm{max}}\left(\left[\begin{array}{cc}
\mathbf{I}\otimes\mathbf{B}_{k} & \mathbf{0}\\
\mathbf{0} & \mathbf{I}\otimes\mathbf{B}_{k}^{\mathrm{T}}
\end{array}\right]\right)-\mu_{g}\sum_{k\in\mathcal{K}_{g}}g_{k}(\gamma)\lambda_{\mathrm{max}}\left(\left[\begin{array}{c}
\mathbf{q}_{k}\\
\mathbf{q}_{k}^{\mathrm{*}}
\end{array}\right]\left[\begin{array}{c}
\mathbf{q}_{k}\\
\mathbf{q}_{k}^{\mathrm{*}}
\end{array}\right]^{\mathrm{H}}\right)\nonumber \\
 & ~~~~~+\lambda_{\mathrm{min}}\left(\mu_{g}\left[\begin{array}{c}
\sum_{k\in\mathcal{K}_{g}}g_{k}(\gamma)\mathbf{q}_{k}\\
\sum_{k\in\mathcal{K}_{g}}g_{k}(\gamma)\mathbf{q}_{k}^{\mathrm{*}}
\end{array}\right]\left[\begin{array}{c}
\sum_{k\in\mathcal{K}_{g}}g_{k}(\gamma)\mathbf{q}_{k}\\
\sum_{k\in\mathcal{K}_{g}}g_{k}(\gamma)\mathbf{q}_{k}^{\mathrm{*}}
\end{array}\right]^{\mathrm{H}}\right)\nonumber \\
 & \overset{\mathrm{(B2)}}{=}-\sum_{k\in\mathcal{K}_{g}}g_{k}(\gamma)\lambda_{\mathrm{max}}(\mathbf{B}_{k})-2\mu_{g}\sum_{k\in\mathcal{K}_{g}}g_{k}(\gamma)\mathbf{q}_{k}^{\mathrm{H}}\mathbf{q}_{k}\nonumber \\
 & \overset{\mathrm{(B2)}}{=}-\sum_{k\in\mathcal{K}_{g}}b_{k}g_{k}(\gamma)\mathbf{e}^{\mathrm{H}}\mathbf{H}_{k}\mathbf{H}_{k}^{\mathrm{H}}\mathbf{e}-2\mu_{g}\sum_{k\in\mathcal{K}_{g}}g_{k}(\gamma)\mathbf{q}_{k}^{\mathrm{H}}\mathbf{q}_{k}\nonumber \\
 & \overset{\mathrm{(B3)}}{\geq}-\mathrm{max}_{k\in\mathcal{K}_{g}}\left\{ b_{k}\mathbf{e}^{\mathrm{H}}\mathbf{H}_{k}\mathbf{H}_{k}^{\mathrm{H}}\mathbf{e}\right\} -2\mu_{g}\mathrm{max}_{k\in\mathcal{K}_{g}}\left\{ ||\mathbf{q}_{k}||_{2}^{2}\right\} \nonumber \\
 & =-\mathrm{max}_{k\in\mathcal{K}_{g}}\left\{ b_{k}\mathbf{e}^{\mathrm{H}}\mathbf{H}_{k}\mathbf{H}_{k}^{\mathrm{H}}\mathbf{e}\right\} -2\mu_{g}\mathrm{max}_{k\in\mathcal{K}_{g}}\left\{ ||\mathbf{Q}_{k}||_{F}^{2}\right\} .\label{eq:eig-1}
\end{align}
\hrule 
\end{figure*}

Recall that $\mathbf{F}=\mathbf{F}^{n}+\gamma(\widetilde{\mathbf{F}}-\mathbf{F}^{n}),\forall\gamma\in[0,1]$,
therefore $||\mathbf{F}^{n}+\gamma(\widetilde{\mathbf{F}}-\mathbf{F}^{n})||_{F}^{2}\leq P_{\mathrm{T}}$.
By using (A4), the last term in the right hand side of the last equation
of (\ref{eq:eig-1}) satisfies inequality (\ref{eq:eig-2}) as 
\begin{align}
||\mathbf{Q}_{k}||_{F}^{2} & =||\mathbf{C}_{k}-\mathbf{B}_{k}^{\mathrm{H}}(\mathbf{F}^{n}+\gamma(\widetilde{\mathbf{F}}-\mathbf{F}^{n}))||_{F}^{2}\nonumber \\
 & =||(\mathbf{F}^{n}+\gamma(\widetilde{\mathbf{F}}-\mathbf{F}^{n}))^{\mathrm{H}}\mathbf{B}_{k}||_{F}^{2}+||\mathbf{C}_{k}||_{F}^{2}\nonumber \\
 & \thinspace\thinspace\thinspace\thinspace\thinspace\thinspace-2\textrm{\ensuremath{\mathrm{Re}}}\left\{ \mathrm{Tr}\left[\mathbf{C}_{k}^{\mathrm{H}}\mathbf{B}_{k}^{\mathrm{H}}(\mathbf{F}^{n}+\gamma(\widetilde{\mathbf{F}}-\mathbf{F}^{n}))\right]\right\} \nonumber \\
 & \overset{\mathrm{(B4)}}{\leq}\lambda_{\mathrm{max}}(\mathbf{B}_{k}\mathbf{B}_{k}^{\mathrm{H}})||\mathbf{F}^{n}+\gamma(\widetilde{\mathbf{F}}-\mathbf{F}^{n})||_{F}^{2}+||\mathbf{C}_{k}||_{F}^{2}\nonumber \\
 & \thinspace\thinspace\thinspace\thinspace\thinspace\thinspace\thinspace-2\textrm{\ensuremath{\mathrm{Re}}}\left\{ \mathrm{\mathrm{Tr}}\left[\mathbf{C}_{k}^{\mathrm{H}}\mathbf{B}_{k}^{\mathrm{H}}(\mathbf{F}^{n}+\gamma(\widetilde{\mathbf{F}}-\mathbf{F}^{n}))\right]\right\} \nonumber \\
 & \leq P_{\mathrm{T}}\lambda_{\mathrm{max}}(\mathbf{B}_{k}\mathbf{B}_{k}^{\mathrm{H}})+||\mathbf{C}_{k}||_{F}^{2}+2\sqrt{P_{\mathrm{T}}}||\mathbf{B}_{k}\mathbf{C}_{k}||_{F}\nonumber \\
 & =P_{\mathrm{T}}b_{k}^{2}|\mathbf{e}^{\mathrm{H}}\mathbf{H}_{k}\mathbf{H}_{k}^{\mathrm{H}}\mathbf{e}|^{2}+||\mathbf{C}_{k}||_{F}^{2}+2\sqrt{P_{\mathrm{T}}}||\mathbf{B}_{k}\mathbf{C}_{k}||_{F}.\label{eq:eig-2}
\end{align}
The third term in the right hand side of the last inequality of (\ref{eq:eig-2})
is the optimal objective value of the following Problem (\ref{eq:Problem-x})
which has a closed-form solution. 
\begin{align}
\mathop{\mathrm{min}}\limits _{\mathbf{X}} & \;\;2\textrm{Re}\left\{ \mathrm{\mathrm{Tr}}\left[\mathbf{C}_{k}^{\mathrm{H}}\mathbf{B}_{k}^{\mathrm{H}}\mathbf{X}\right]\right\} \nonumber \\
{\rm s.t.} & \thinspace\thinspace\thinspace\thinspace\mathrm{\mathrm{Tr}}\left[\mathbf{X}^{\mathrm{H}}\mathbf{X}\right]\leq P_{\mathrm{T}}.\label{eq:Problem-x}
\end{align}

Finally, combining (\ref{eq:eig-1}) with (\ref{eq:eig-2}), we arrive
at (\ref{alpha}). Hence, the proof is complete.

\section{The proof of Theorem \ref{Theorem-2}\label{subsec:The-proof-of-4}}

Since $f_{g}\left(\mathbf{e}\right)$ is twice differentiable and
concave, we minorize $f_{g}\left(\mathbf{e}\right)$ at $\mathbf{e}^{n}$
with a quadratic function, as follows 
\begin{align}
f_{g}(\mathbf{e})\geq & f_{g}(\mathbf{e}^{n})+2\textrm{\ensuremath{\mathrm{Re}}}\left\{ \mathbf{d}_{g}^{\mathrm{H}}(\mathbf{e}-\mathbf{e}^{n})\right\} \nonumber \\
 & +(\mathbf{e}-\mathbf{e}^{n})^{\mathrm{H}}\mathbf{N}_{g}(\mathbf{e}-\mathbf{e}^{n}),\label{eq:quadratic-e}
\end{align}
where vectors $\mathbf{d}_{g}\in\mathbb{C}^{M\times1}$ and matrices
$\mathbf{N}_{g}\in\mathbb{C}^{M\times M}$ are determined to satisfy
conditions (A1)-(A4).

Obviously, (A1) and (A4) are already satisfied. In order to satisfy
condition (A3), the directional derivatives of $f_{g}(\mathbf{e})$
and the right hand side of (\ref{eq:quadratic-e}) must be equal,
yielding 
\begin{equation}
\mathbf{d}_{g}=\sum_{k\in\mathcal{K}_{g}}g_{k}(\mathbf{e}^{n})(\mathbf{a}_{k}-\mathbf{A}_{k}^{\mathrm{H}}\mathbf{e}^{n}),
\end{equation}
where $g_{k}(\mathbf{e}^{n})$ is defined in (\ref{eq:gke}).

Let $\mathbf{e}=\mathbf{e}^{n}+\gamma(\widetilde{\mathbf{e}}-\mathbf{e}^{n}),\forall\gamma\in[0,1]$.
In order to satisfy condition (A2), it suffices to show 
\begin{align}
f_{g}(\mathbf{e}^{n}+\gamma(\widetilde{\mathbf{e}}-\mathbf{e}^{n})) & \geq f_{g}(\mathbf{e}^{n})+2\gamma\textrm{\ensuremath{\mathrm{Re}}}\left\{ \mathbf{d}_{g}^{\mathrm{H}}(\widetilde{\mathbf{e}}-\mathbf{e}^{n})\right\} \nonumber \\
 & +\gamma^{2}(\widetilde{\mathbf{e}}-\mathbf{e}^{n})^{\mathrm{H}}\mathbf{N}_{g}(\widetilde{\mathbf{e}}-\mathbf{e}^{n}).\label{eq:quadratic-e-A2}
\end{align}
Then, we need to calculate the second-order derivatives of the left
hand side and the right hand side of (\ref{eq:quadratic-e-A2}), and
make the latter one lower than or equal to the former for $\forall\gamma\in[0,1]$
and $\forall\widetilde{\mathbf{e}},\forall\mathbf{e}^{n}\in\mathcal{S}_{e}$.

The second-order derivative of the left hand side of (\ref{eq:quadratic-e-A2})
is given by 
\begin{align}
\frac{\partial^{2}L_{g}(\gamma)}{\partial\gamma^{2}} & =\left[\begin{array}{cc}
\mathbf{t}{}^{\mathrm{H}} & \mathbf{t}{}^{\mathrm{T}}\end{array}\right]\boldsymbol{\Psi}_{g}\left[\begin{array}{c}
\mathbf{t}\\
\mathbf{t}{}^{*}
\end{array}\right],\label{eq:second-left-e}
\end{align}
with $\mathbf{t}=\widetilde{\mathbf{e}}-\mathbf{e}^{n}$. $\boldsymbol{\Psi}_{g}$
is shown in (\ref{Psi}) where 
\begin{figure*}
\begin{align}
\boldsymbol{\Psi}_{g}=\sum_{k\in\mathcal{K}_{g}}\left(g_{k}(\gamma)\left[\begin{array}{cc}
-\mathbf{A}_{k} & 0\\
\mathbf{0} & -\mathbf{A}_{k}^{\mathrm{T}}
\end{array}\right]-\mu_{g}g_{k}(\gamma)\left[\begin{array}{c}
\mathbf{q}_{k}\\
\mathbf{q}_{k}^{\mathrm{*}}
\end{array}\right]\left[\begin{array}{c}
\mathbf{q}_{k}\\
\mathbf{q}_{k}^{\mathrm{*}}
\end{array}\right]^{\mathrm{H}}\right)+\mu_{g}\left[\begin{array}{c}
\sum_{k\in\mathcal{K}_{g}}g_{k}(\gamma)\mathbf{q}_{k}\\
\sum_{k\in\mathcal{K}_{g}}g_{k}(\gamma)\mathbf{q}_{k}^{\mathrm{*}}
\end{array}\right]\left[\begin{array}{c}
\sum_{k\in\mathcal{K}_{g}}g_{k}(\gamma)\mathbf{q}_{k}\\
\sum_{k\in\mathcal{K}_{g}}g_{k}(\gamma)\mathbf{q}_{k}^{\mathrm{*}}
\end{array}\right]^{\mathrm{H}},\label{Psi}
\end{align}
\hrule 
\end{figure*}

\begin{align}
\mathbf{q}_{k} & =\mathbf{a}_{k}-\mathbf{A}_{k}^{\mathrm{H}}(\mathbf{e}^{n}+\gamma(\widetilde{\mathbf{e}}-\mathbf{e}^{n}))\\
g_{k}(\gamma) & =\frac{\mathrm{exp}\left\{ -\mu_{g}l_{k}(\gamma)\right\} }{\sum_{k\in\mathcal{K}_{g}}\mathrm{exp}\left\{ -\mu_{g}l_{k}(\gamma)\right\} },k\in\mathcal{K}_{g}
\end{align}

The second-order derivative of the right hand side of (\ref{eq:quadratic-e-A2})
is 
\begin{align}
 & 2(\widetilde{\mathbf{e}}-\mathbf{e}^{n})^{\mathrm{H}}\mathbf{N}_{g}(\widetilde{\mathbf{e}}-\mathbf{e}^{n})\nonumber \\
= & \left[\begin{array}{cc}
\mathbf{t}{}^{\mathrm{H}} & \mathbf{t}{}^{\mathrm{T}}\end{array}\right]\left[\begin{array}{cc}
\mathbf{I}\otimes\mathbf{N}_{g} & \mathbf{0}\\
\mathbf{0} & \mathbf{I}\otimes\mathbf{N}_{g}^{\mathrm{T}}
\end{array}\right]\left[\begin{array}{c}
\mathbf{t}\\
\mathbf{t}{}^{*}
\end{array}\right].\label{eq:second-right-e}
\end{align}

Combining (\ref{eq:second-left-e}) with (\ref{eq:second-right-e}),
$\mathbf{N}_{g}$ must satisfy 
\[
\boldsymbol{\Psi}_{g}\succeq\left[\begin{array}{cc}
\mathbf{I}\otimes\mathbf{N}_{g} & \mathbf{0}\\
\mathbf{0} & \mathbf{I}\otimes\mathbf{N}_{g}^{\mathrm{T}}
\end{array}\right].
\]
For simplicity, we choose $\mathbf{N}_{g}=\beta_{g}\mathbf{I}=\lambda_{\mathrm{min}}(\boldsymbol{\Psi}_{g})\mathbf{I}$.
Eventually, (\ref{eq:quadratic-e}) is equivalent to 
\begin{align}
f_{g}(\mathbf{e}) & \geq f_{g}(\mathbf{e}^{n})+2\textrm{\ensuremath{\mathrm{Re}}}\left\{ \mathbf{d}_{g}^{\mathrm{H}}(\mathbf{e}-\mathbf{e}^{n})\right\} +\beta_{g}(\mathbf{e}-\mathbf{e}^{n})^{\mathrm{H}}(\mathbf{e}-\mathbf{e}^{n})\nonumber \\
 & =2\textrm{\ensuremath{\mathrm{Re}}}\left\{ \mathbf{u}_{g}^{\mathrm{H}}\mathbf{e}\right\} +\textrm{consE}_{g},\label{eq:surrogate-e-final}
\end{align}
where $\mathbf{u}_{g}$, $\beta_{g}$, and $\textrm{consE}_{g}$ are
given in (\ref{u}), (\ref{beta}), and (\ref{eq:constant-e}), respectively.
The last equation of (\ref{eq:surrogate-e-final}) is from the unit-modulus
constraints, i.e., $\mathbf{e}^{\mathrm{H}}\mathbf{e}=(\mathbf{e}^{n})^{\mathrm{H}}\mathbf{e}^{n}=M+1$.
The method to get the value of $\beta_{g}$ is similar as $\alpha_{g}$,
so we omit it here. Hence, the proof is complete.

\section{The proof of Theorem \ref{Theorem-MM-kkt}\label{subsec:The-proof-of-6}}

Let us denote the converged solution of Problem (\ref{eq:Problem-2})
by $\{\mathbf{F}^{o},\mathbf{e}^{o}\}$. In the following, we prove
that $\{\mathbf{F}^{o},\mathbf{e}^{o}\}$ satisfies the KKT conditions
of Problem (\ref{eq:Problem-2}).

Firstly, since $\mathbf{F}^{o}$ is the globally optimal solution
of Problem (\ref{eq:Problem-f-final}), the KKT conditions of the
Lagrangian in (\ref{eq:L-mm-F}) of Problem (\ref{eq:Problem-f-final})
is given by 
\begin{align}
 & \sum_{g=1}^{G}\nabla_{\mathbf{F}^{*}}\widetilde{f}_{g}(\mathbf{F}|\mathbf{F}^{n})|_{\mathbf{F}=\mathbf{F}^{o}}-\tau^{o}\mathbf{F}^{o}=\mathbf{0},\label{eq:KKT-mmf-1}\\
 & \tau^{o}(\mathrm{Tr}\left[\mathbf{F}^{\mathrm{H},o}\mathbf{F}^{o}\right]-P_{\mathrm{T}})=0,\label{eq:KKT-mmf-2}
\end{align}
where $\tau^{o}$ is the optimal Lagrange multiplier. According to
the condition (A3), we have 
\begin{align}
\nabla_{\mathbf{F}^{*}}\widetilde{f}_{g}(\mathbf{F}|\mathbf{F}^{n})|_{\mathbf{F}=\mathbf{F}^{o}} & =\nabla_{\mathbf{F}^{*}}f_{g}\left(\mathbf{F},\mathbf{e}^{o}\right)|_{\mathbf{F}=\mathbf{F}^{o}}.\label{eq:A3-f-1}
\end{align}
By substituting (\ref{eq:A3-f-1}) into (\ref{eq:KKT-mmf-1}), we
arrive at 
\begin{align}
 & \sum_{g=1}^{G}\nabla_{\mathbf{F}^{*}}f_{g}\left(\mathbf{F},\mathbf{e}^{o}\right)|_{\mathbf{F}=\mathbf{F}^{o}}-\tau^{o}\mathbf{F}^{o}=\mathbf{0},\label{eq:KKT-mmf-3}
\end{align}

Then, since $\mathbf{e}^{o}$ is the locally optimal solution of Problem
(\ref{eq:Problem-e-final}), it is readily to obtain the following
KKT conditions: 
\begin{align}
 & \sum_{g=1}^{G}\nabla_{\mathbf{e}^{*}}f_{g}\left(\mathbf{F}^{o},\mathbf{e}\right)|_{\mathbf{e}=\mathbf{e}^{o}}-\sum_{m=1}^{M}\tau_{m}^{(2),o}(\nabla_{\mathbf{e}^{*}}|e_{m}|)|_{\mathbf{e}=\mathbf{e}^{o}}\nonumber \\
 & -\tau_{M+1}^{(2),o}(\nabla_{\mathbf{e}^{*}}e_{M+1})|_{\mathbf{e}=\mathbf{e}^{o}}=\mathbf{0},\label{eq:KKT-mme-1}\\
 & \tau_{m}^{(2),o}(|e_{m}^{o}|-1)=0,1\leq m\leq M,\tau_{M+1}^{(2),o}(e_{M+1}^{o}-1)=0,\label{eq:KKT-mme-2}
\end{align}
where $\boldsymbol{\tau}^{(2),o}=[\tau_{1}^{(2),o},...,\tau_{M+1}^{(2),o}]$
are the optimal Lagrange multipliers.

Then, the set of equations (\ref{eq:KKT-mmf-3}), (\ref{eq:KKT-mmf-2}),
(\ref{eq:KKT-mme-1}), and (\ref{eq:KKT-mme-2}) constitute exactly
the KKT conditions of Problem (\ref{eq:Problem-2}).

Hence, the proof is complete.

 \bibliographystyle{IEEEtran}
\bibliography{bibfile}

\end{document}